\documentclass[single]{aastex62}

\usepackage{array}
\usepackage{multirow}
\usepackage[utf8]{inputenc}

\graphicspath{{./}{figures/}}

\newcommand{\beq}{\begin{equation}}
\newcommand{\eeq}{\end{equation}}

\shorttitle{Observations with SCSI. I.}
\shortauthors{Horch et al.}

\begin{document}
\title{Observations with the Southern Connecticut Stellar Interferometer. 
I. Instrument Description and First Results}

\correspondingauthor{Elliott Horch}
\email{horche2@southernct.edu}

\author{Elliott P. Horch}
\altaffiliation{Adjunct Astronomer, Lowell Observatory}
\affiliation{Department of Physics, Southern Connecticut State University, 501 Crescent Street, 
New Haven, CT 06515, USA}

\author{Samuel A. Weiss}
\affiliation{Department of Physics, Southern Connecticut State University, 501 Crescent Street, 
New Haven, CT 06515, USA}

\author{Paul M. Klaucke}
\affiliation{Department of Physics, Southern Connecticut State University, 501 Crescent Street, 
New Haven, CT 06515, USA}

\author{Richard A. Pellegrino}
\affiliation{Department of Physics, Southern Connecticut State University, 501 Crescent Street, 
New Haven, CT 06515, USA}

\author{Justin D. Rupert}
\affiliation{MDM Observatory, c/o Kitt Peak National Observatory, 950 North Cherry Avenue, Tucson,
AZ 85719, USA}





\begin{abstract}
We discuss the design, construction, and operation of a new intensity interferometer,
based on the campus of Southern Connecticut State University in New Haven, Connecticut.
While this paper will focus on observations taken with an original two-telescope configuration, 
the current instrumentation consists of three portable 0.6-m Dobsonian telescopes
with single-photon avalanche diode (SPAD) detectors located at the Newtonian 
focus of each telescope. Photons detected at each station are time-stamped and
read out with timing correlators that can give cross-correlations in timing to a precision of 48 ps. 
We detail our observations  to date with the system,
which has now been successfully used at our university in 16 nights of observing.
Components of the instrument were also 
deployed on one occasion at Lowell Observatory, where the Perkins and Hall telescopes
were made to function as an intensity interferometer. We characterize the performance
of the instrument in detail. In total, the observations indicate the detection of a correlation
peak at the level of 6.76$\sigma$ when observing unresolved stars, and consistency with partial or no
detection when observing at a baseline sufficient to resolve the star. Using these measurements
we conclude that the angular diameter of Arcturus is larger than 15 mas, and that of 
Vega is between 0.8 and 17 mas. While the uncertainties
are large at this point, both results are consistent 
with measures from 
amplitude-based long baseline optical interferometers.
\end{abstract}

\keywords{Astronomical Instrumentation --- Interferometers --- Field Stars --- Stellar Radii
}

\section{Introduction}

Intensity interferometry, first proposed by \citet{han56,han58}, can be
used as a method of 
extracting very high resolution information of astrophysical sources based on the super-Gaussian
statistics of photons, also referred to as the Hanbury Brown and Twiss (HBT) effect. They showed
that light recorded by two independent detectors at different locations has a weak correlation in 
intensity if the object being viewed was indistinguishable from a  point source. In astronomical terms,
this implies that if the two detectors are separated by a distance less than that needed to resolve
a star, then the intensity correlation should be observed. In contrast, when the distance 
is increased to the
point where the baseline resolves the object, the correlation disappears. In between these two 
limits, partial correlation can be observed that traces out the profile of an Airy pattern, the width of 
which is inversely proportional to the diameter of the star being observed. Hanbury Brown and
his collaborators were able to measure the diameters of 32 stars from the site of 
their famous Narrabri intensity interferometer \citep{han74}, which represented the culmination of their 
intensity interferometry efforts. Chronologically, this work sat between the seminal interferometric 
studies of Michelson and Pease 
performed at the Mount Wilson 2.5-m 
telescope \citep{mic21,pea25} and the development and
use of the Mark III stellar interferometer in the 1980's and 1990's \citep{sho88,hut89,hum95}. 
The latter paved the way for the return to amplitude-based
optical interferometry and the rise of specialized facilities for long baseline optical interferometry such as SUSI 
\citep{dav94,tan03,dav05},  
the Navy Precision Optical Interferometer \citep{arm98,nor99,bai18}, 
the CHARA Array \citep{mca05, ten05, sch20},
and the Magdalena Ridge Optical Interferometer \citep{cre04, cre18, cre20}.

Starting in the mid-2000's, however, a revival of intensity interferometry has slowly taken shape. 
One of the principal reasons for this is that ultra-fast detectors with high quantum efficiencies have
become widely available. This allows for a re-examination of the technique in a photon-counting
context with the advantage that increased speed creates the possibility of higher signal-to-noise
ratios (SNRs) as detailed in \citet{kle07}.
Those authors discuss the fact that the timing capabilities of modern detectors and 
timing correlators can be as much as a 
factor of 1000 higher than those of the instrumentation used at Narrabri, which gives flexibility
in the design of a modern intensity interferometer. For example, whereas the Narrabri instrument 
used two telescopes with primary mirror diameters of 6.5 m to achieve a limiting magnitude of 
approximately $V$=2, the higher timing resolution of modern detectors allows one to achieve the
same limiting magnitude with telescopes that are much smaller. Alternatively, significantly fainter 
limiting magnitudes are now possible with telescopes that are comparable in size to the Narrabri
instrument. Several groups have 
built new intensity interferometers and reported successful correlation measurements in recent 
years, including \citet{riv18, riv20, acc20, abe20}; and \citet{zam21}. The intrinsic timing 
resolution of these instruments
spans a range from roughly 400 ps to 2.5 ns and they use telescopes with diameters of 1 to 17 m;
in the case of larger telescopes \citep{acc20, abe20}, 
these are so-called ``light-bucket'' telescopes, which do not
have high optical resolution, and as a consequence, detectors with a larger active area must be 
used to collect the light, generally photomultiplier tubes. 
Thus these instruments are much closer in design to the Narrabri stellar
interferometer of the 1970s. In contrast, the employment of smaller ($\sim$1-m class) telescopes
\citep{riv18, zam21} has generally been coupled with the use of ultra-fast single-photon 
avalanche diode (SPAD) detectors.

Over the last several years, our group has designed, built, and operated a new intensity 
interferometer based mainly at the campus of Southern Connecticut State University, but 
which is modular enough to transport the photon-detection instrumentation to other astronomical
facilities \citep{hor12,hor16,hor18,wei18,kla20}. 
It belongs to the second category of intensity interferometers just mentioned in that it uses
0.6-m telescopes and SPAD detectors.
We describe in this paper the final instrument design and construction, 
discuss the observations taken to date both on 
our campus and at the Anderson Mesa site at Lowell Observatory, and analyze 
 the results obtained with
the data so far. While we are still improving the instrument performance and the data-taking process,
these observations together indicate that the system is detecting 
intensity correlations at the level predicted by theory. Our instrument uses smaller telescopes
than any of the other intensity interferometers mentioned above and its timing precision is 
higher, thus it pushes the technique into a regime where it can be done with portable telescopes
and at relatively low cost.
We use the correlations detected to make deductions concerning the diameters of Vega ($\alpha$ Lyr, 
HR 7001), 
and Arcturus ($\alpha$ Boo, HR 5340);
while still uncertain, these measurements are consistent with those made with the larger 
Michelson-style optical interferometers. 

\section{Basic Theory}

The details of optical detection depend on the properties of photons,
which follow Bose-Einstein statistics. As discussed in e.g.\ \citet{man63} and \citet{han74b}, if $n$ 
photons are detected on average in a time interval $\Delta t$, then the variance in the detection 
rate for a linearly polarized input beam is given by

\beq
(\Delta n)^2 = n + n^2 \frac{\Delta \tau}{\Delta t},
\eeq

\noindent
where $\Delta \tau$ represents the coherence timescale over which intensity variations
occur due to Bose-Einstein fluctuations and is related to the inverse of the frequency
bandwidth of the detected light. The presence of the second term on the right-hand side of 
the equation indicates that over time intervals of order $\Delta \tau$ or less there 
will be an increased 
probability of photon detection, which represents the HBT effect and is sometimes
referred to as photon bunching. In most 
applications, however, $\Delta \tau$ is at least several orders of magnitude smaller than $\Delta t$ 
and the deviation from Poisson statistics is negligible.

The possibility for using this effect in high-resolution imaging in astronomy arises from the 
fact that the second order coherence function, which depends on the expression
for the variance above, is also related to the modulus
square of the complex visibility. The former is defined as

\beq
g^{(2)}({\bf r},{\bf b},t,t_0) = \frac{\left< I_1({\bf r},t) \cdot I_2({\bf r}+{\bf b},t+t_0) \right>}{\left<I_1({\bf r},t)\right> \cdot \left<I_2({\bf r},t)\right>},
\eeq

 \noindent
 where the brackets indicate a time average, $I_1$ and $I_2$ are the irradiance values
 at two different detection points, ${\bf r}$ is a fixed detection point in space,
 ${\bf b}$ is the baseline between ${\bf r}$ and a second detection location, and $t_{0}$ is the
 timing delay between detection at the two stations. 
 Defining the complex visibility as $V_{12}$, the relationship between 
 $g^{(2)}$ and $V_{12}$ in the case of linearly polarized light can be written as
 
 \beq
 g^{(2)}_{\rm pol} = 1 +  |V_{12}|^2 \frac{\Delta \tau}{\Delta t}.
 \eeq
 
 \noindent
For the case of unpolarized light, \citet{man63} shows that, because the second order coherence
function involves the products of intensities and not field amplitudes,
the expected result for $g^{(2)}$ for unpolarized light modifies the 
last term in the above by a factor of 1/2:
 
  \beq
 g^{(2)}_{\rm unpol} = 1 + \frac{1}{2} |V_{12}|^2 \frac{\Delta \tau}{\Delta t}.
 \eeq
 
 \noindent
 Regardless of the polarization state of the detected light, Equations 3 and 4
 establish the fact that high-resolution information about a source 
can be obtained by detecting the irradiance at two locations without physically
interfering the light. For a photon-counting experiment, Equation 2 indicates that
the data product needed to obtain $g^{(2)}$ (and therefore $|V_{12}|^{2}$) is the
normalized timing cross-correlation
of the photon data streams.
However, in order to retrieve high-quality data and determine the
visibility, it is crucial to increase the fraction $\Delta \tau / \Delta t$ as much as possible.
This can be done by decreasing the frequency bandwidth of the light by using an extremely
narrow-bandpass filter (which increases the numerator), or by increasing the
electronic bandwidth of the detector (which decreases the denominator), or both. Much of 
the recent revival of intensity interferometry has been made possible by detector
developments that have enabled the second of these options.

To estimate the signal-to-noise ratio (SNR) that is practically obtainable
in the unpolarized case, we follow \citet{mal14}. 
In a given timing interval $\Delta t$, there will be approximately  $\Delta t / \Delta \tau$ intervals of
coherence. If the count rate per second is given by $r$, then the number of counts 
detected per interval at each telescope is given by $n = r \Delta t$. The 
total number of photon correlations generated per timing 
interval is the square of this, 
$(r \Delta t)^2$, and each of these has a probability of 
$\Delta \tau / \Delta t$ of being correlated due to the HBT effect 
(with the remainder being ``random'' correlations), yielding a signal of 
\beq
S = \frac{1}{2} (r \Delta t)^2 \cdot \frac{\Delta \tau}{\Delta t} = \frac{r^2 \Delta \tau \Delta t.}{2}
\eeq

\noindent
On the other hand, the noise would be dominated by Poisson statistics even at
the detector speeds available today, and 
given by the square root of the number of random correlations 
obtained per interval, or in other words,
\beq
N = r \Delta t.
\eeq

\noindent
Therefore, the SNR is given by
\beq
{\rm SNR}(\Delta t) = \frac{r \Delta \tau}{2}.
\eeq

\noindent
Over a longer integration time, the result from each counting interval would add in quadrature,
so that in a total exposure time of $T$, the final SNR would be given by
\beq
{\rm SNR}(T) = \frac{r \Delta \tau}{2} \sqrt{\frac{T}{\Delta t}}.
\eeq

\noindent
All of the quantities on the right-hand side of this equation are direct observables
except the coherence time $\Delta \tau$. However, 
this is easily obtained from characteristics of the filter used to make the observations. We 
calculate the frequency bandwidth from filter properties as
\beq 
\Delta \nu = \frac{c}{\lambda_{0} - \Delta \lambda / 2} - \frac{c}{\lambda_{0}+\Delta \lambda / 2}, 
\eeq

\noindent
where $\lambda_{0}$ is the center wavelength of the observation and $\Delta \lambda$ is the 
full width at half maximum (FWHM) of the filter transmission. While the coherence time
depends to a certain extent on the shape of the filter transmission curve, it can  
be approximated as


\beq
\Delta \tau \approx \frac{1}{\Delta \nu}.
\eeq
\noindent
Once $\Delta \tau$ is calculated for a given filter, the fraction of photons correlated due to
the HBT effect is $\Delta \tau / (2 \Delta t)$ in the unpolarized case. For 
example, with a filter width of 3 nm and center wavelength of 532 nm,
 $\Delta \tau \approx 0.31$ ps; using a timing resolution of 64 ps to record the data, the 
expected fraction of correlated photons is then 0.0024. At a constant count rate of 1 MHz at each
telescope (which, as we will show in Section 4, we can generally obtain on the 
2nd magnitude star Polaris [$\alpha$ UMi]), 
a 5-$\sigma$ correlation peak above the noise (SNR = 5) would be obtained 
when $T \approx 19$ hours of observing an unresolved star ($|V_{12}|^2 = 1$). 
Using a narrower filter would result in a larger fraction
of correlated photons but fewer photons detected in the same proportion, so the SNR for 
a given observing time $T$ remains the same.

\section{Instrument Description}

The instrument we have constructed and used to date
is a two-station interferometer
where, for on-campus observing,
the two telescopes may be easily moved to observe at a desired
baseline and orientation. The telescopes used are 0.6-m Dobsonian
telescopes made by Equatorial Platforms of Grass Valley, California.
Each has a focal length of 2 m, 
so that the plate scale at the Newtonian focus is approximately 0.1 arcsec per 
micron. 
We have recently obtained a third identical telescope, and future observations
with the instrument will be made with all three stations deployed. 

An image of the main components and a block diagram
of the interferometer are shown in Figure 1.
We place a single photon avalanche diode (SPAD) detector at the Newtonian focus
of each telescope. Until 2019, we used two Micro Photon Devices (MPD) PDM series 
SPADs with a 50 micron-diameter active area; given the plate scale mentioned above,
this maps to a 5-arcsecond diameter region on the sky. Starting in 2019, we
began using two 100-micron diameter PDM-series SPADs. We also have a
 $1 \times 8$ SPAD array from the
Politecnico di Milano, Italy. A detailed description
of the array is given in \citet{cam12}; however, this device was only 
used on one occasion for the observations described below.
Generally, the timing uncertainty of cross-correlations made with any of these
detectors is about 50 ps,
and the dead time is about $\sim$80 ns after each detected photon. This
limits the count rate that can be detected to less than about 10 MHz,
but for astronomical purposes, this is not a significant limitation
since the expected count rate will be below this limit
for most stars we wish to observe.

\begin{figure}[t]
\figurenum{1}
\includegraphics[scale=1.5]{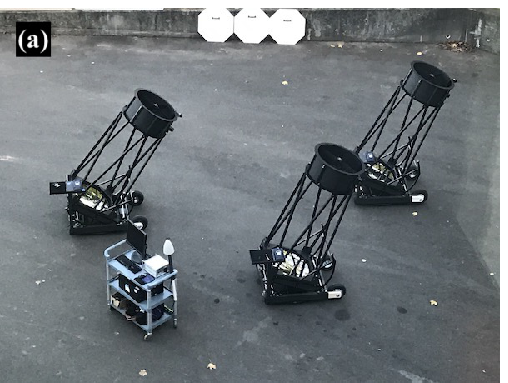}
\includegraphics[scale=0.375]{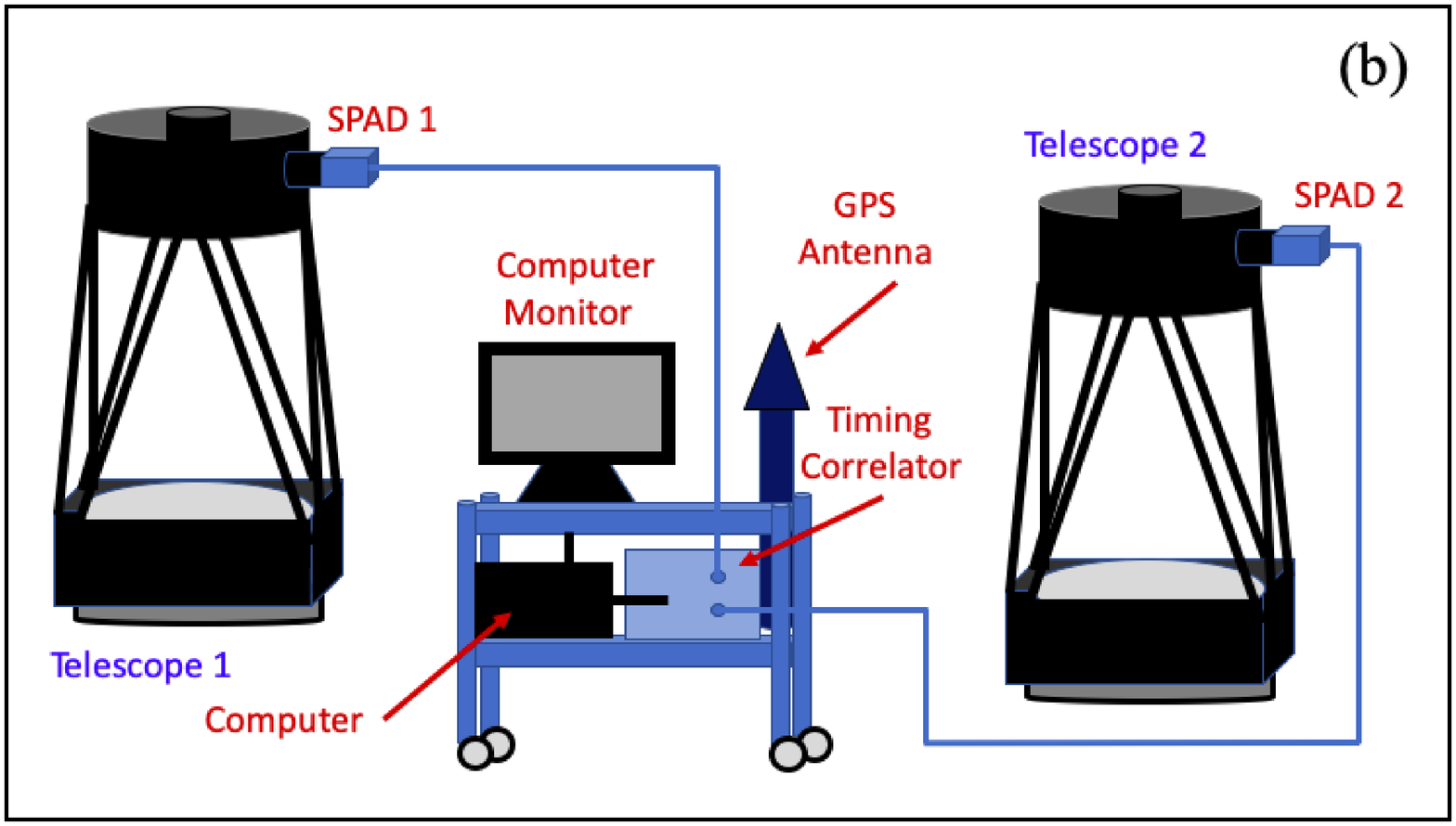}
\caption{
(a) The instrument set up at the standard observing location on the SCSU campus. (b) A block diagram of the simpler two-telescope arrangement, which will be the focus of this paper.}
\end{figure}

To mount the detector on the telescope, it is placed in a harness made from 
a tube of PVC pipe. The tube is cut so that it is in contact with the body of the detector and holds
it from the top and bottom. The arrangement is held in place with elastic bands that create
a pressure fit onto the back of the tube. Inside the tube is an optical assembly
consisting of a 35-mm focal length acromat lens, followed by a narrow band pass
filter, followed by a second lens that is identical to the first. With the front lens at the correct
position in the focus tube of the telescope, a collimated beam is created between
the lenses, as the light passes through the filter. Once reimaged by the second lens,
the light is focused on the detector with no change in plate scale relative to the
Newtonian focus. We currently have three filters available for use with the set up
that have specifications shown in Table 1. No polarizers are used in the optical system.

Two timing correlators have been used with the instrument so far to read out
and record the events detected by the SPAD detectors: a PicoHarp 300
and a HydraHarp 400, both made by PicoQuant, Inc. The former has
two input channels, whereas the latter has eight. Both devices have
a time-tagging mode, where each photon event that is detected in
any channel is time-stamped, and the timing address and channel number
are written out in a 4-byte block. In this mode, the least significant bit
of the address corresponds to 4 ps in the case of the PicoHarp
and 1 ps in the case of the HydraHarp. For the PicoHarp, 
the timing addresses
are the lower 28 bits of this address and the upper four bits encode
the channel number, so that in a typical observation for a single data file
of 10s or 100s
of seconds, the timing address will roll over many times, roughly once per
0.8 ms. Each time
the counter rolls over, an event is written with a special code
in the top 4 bits to indicate that a rollover
has occurred rather than a photon
event. 
By keeping track of the total number of roll-overs that have occurred as the events
are subsequently processed, a unique timing can be associated to each
event even in a long observation. For the HydraHarp, the upper six 
bits of the four byte data word are reserved for encoding the channel number
corresponding to the event and special event markers; in this case, because
the number of timing bits is lower and yet the timing resolution is higher,
rollovers in timing occur roughly once per 30 microseconds. To minimize
the number of rollover markers written when count rates are low, 
the rollover data word also contains
the number of rollovers since the last photon event. 

Events are read out with a small desktop computer that is placed on a table or a 
cart near the observing site. An observation file is stored with a header that contains
basic instrument parameters, followed by the sequence of time stamps for recorded events.
A typical 5-minute file 
where the count rate on each channel is 1 MHz results in a file that is approximately 2.5 GB 
in size; a typical observing session of 4 hours can therefore yield over 100 GB of data. 

\begin{deluxetable}{lccc}
\tablewidth{0pt}
\tablenum{1}
\tablecaption{Current SCSI Filter Options\tablenotemark{a}}
\tablehead{
\colhead{Manufacturer} &
\colhead{$\lambda_{0}$} &
\colhead{$\Delta \lambda$} &
\colhead{Max.} \\
& \colhead{(nm)} &  \colhead{(FWHM, nm)} & \colhead {Trans.}
}
\startdata
Edmund Optics & 532 & 2.02 -- 3.7 & 0.90 \\
Edmund Optics & 532 & 1.2 & 0.95 \\
Newport Optics & 633 & 1.0 & 0.30 \\
\enddata
\tablenotetext{a}{All values are as stated by the filter vendors.}
\end{deluxetable}

\section{Expected System Performance}

While optical path length requirements in intensity interferometry are not as stringent
as in amplitude-base interferometry, our instrument
relies on photon counting with extremely high timing resolution. Given this, 
coincidence of photon arrivals can be affected at a level 
important for our data reduction in three main ways: 
(1) electronic delay based on detector response, propagation of the signals through cable, 
and intrinsic timing differences between readout channels in the timing correlators,
(2) optical delay within each telescope system, and 
(3) geometrical delays based on the location of the star on the sky at the time of 
observation.  
We discuss each aspect of the
performance that we can expect from our instrument below.

\subsection{Electronic Considerations}

Detector response is the primary factor that will influence the detected correlation. 
Our detectors have measured FWHM timing uncertainties of approximately 30 ps. 
This means that two photons that are correlated are not generally 
detected as coincident in time, but could differ by tens of ps in terms of when the first detector pulse is created 
versus the second. The correlation signal, which is a delta function in the perfect case, is therefore
smeared out in time. The timing correlator then time-stamps the pulses as they arrive, and when the
cross-correlation function is computed from the events collected, the delta function is approximately
recovered if the timing interval $\Delta t$ is chosen to match the width of the smeared peak.
In our case, because each detector has a FWHM of 30 ps, the cross-correlation width should be
approximately $30 \cdot \sqrt{2} = $ 42 ps in width, just from the detector timing characteristics. 

Two further contributions to this smearing arise from the spreading of the pulse as it travels 
through the cable to the timing module and from the intrinsic timing resolution of the timing 
correlator itself (regardless of the precision of the timing address). To measure these effects,
 we split the output cable from 
a single detector into two lines, and fed these
into two (independent) channels of the timing correlator in the laboratory. 
We then exposed the detector assembly (the SPAD detector and the fore-optics) to 
diffuse light, and read out pulses from both channels. In this case, the uncertainty in the detector
timing is eliminated because we correlate the same pulses, just repeated in the two channels. In
cross-correlating events from a data file taken in this way, we find a FWHM of approximately 
24 ps for the typical cable lengths that we use for on-campus observing. 
On this basis, we conclude that the final intrinsic limitation of the timing resolution of a 
cross-correlation between channels is approximately $\sqrt{42^2 + 24^2} = 48$ ps 
for our system.

Using the same experimental set-up, we can also measure any intrinsic timing offset between
the read-out channels as well as the timing delay associated with each cable. Figure 2(a) shows
sample results that we obtain when inserting three short extra cable lengths into one of the two 
channels prior to taking data. Repeated measurements indicate that cable delays are determined
with a precision of approximately 6 ps. We also find that small intrinsic offsets do exist between
the channels of our correlators, on the order of a few tens of ps for the two channels in
the PicoHarp, and as much as $\sim$400 ps for channels in the HydraHarp. 
Furthermore, these offsets appear to drift over a period of months by 10-20 ps. To account for
this, we measured the offsets and cable delays about 3-4 times per year, and apply the 
offsets obtained closest in time to the observations being reduced.

\begin{figure}[!t]
\figurenum{2}
\plottwo{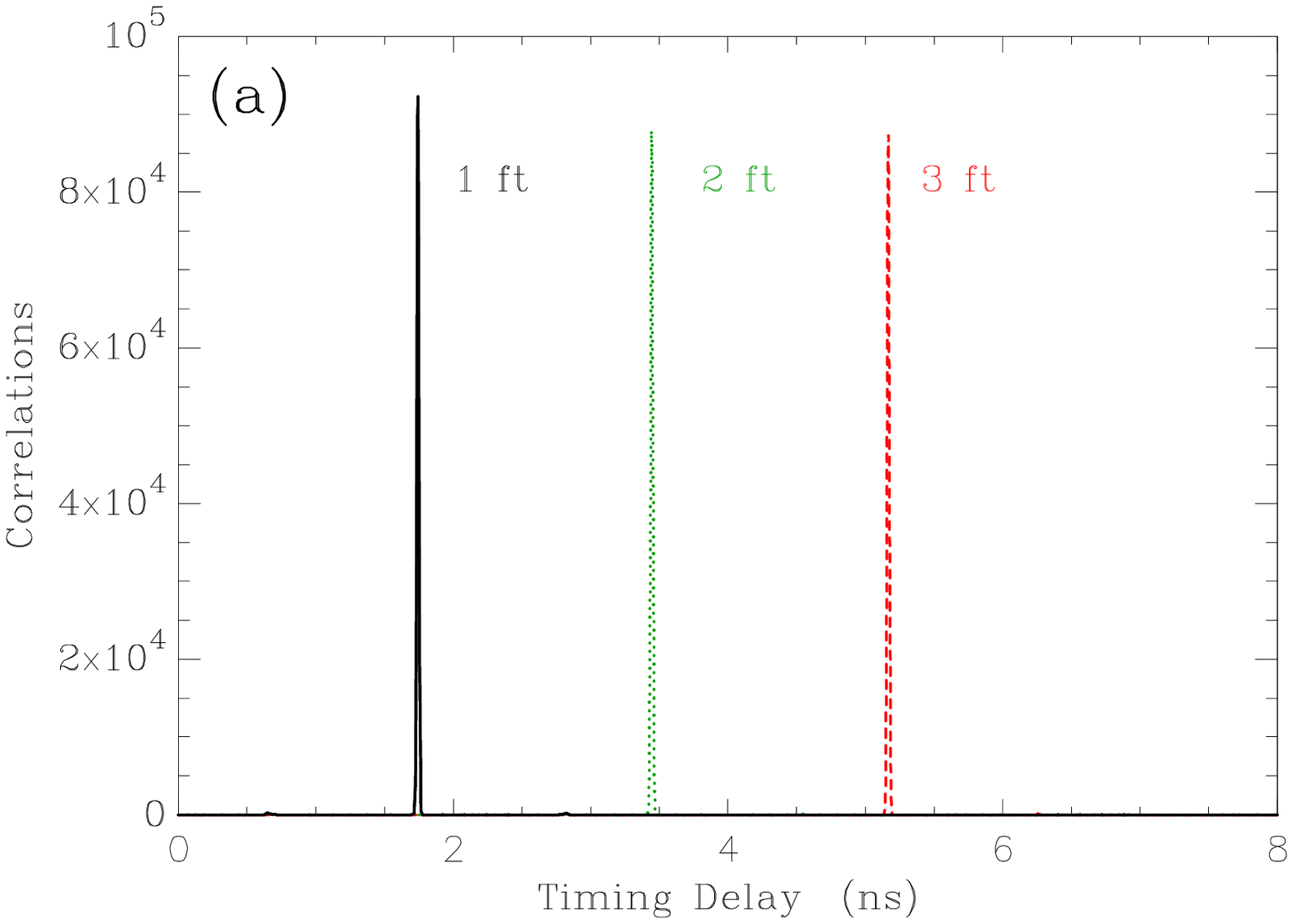}{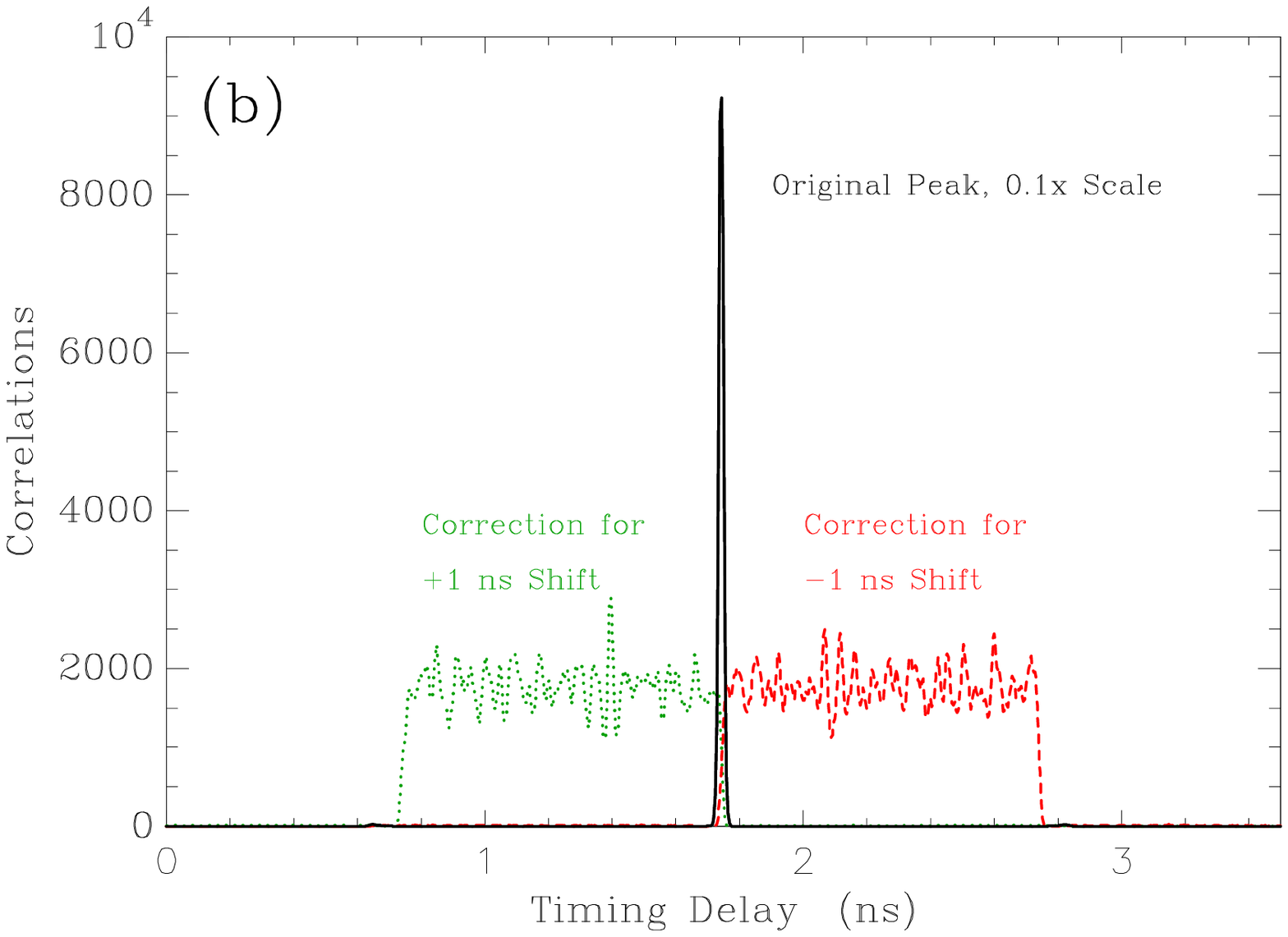}
\figcaption{(a) Correlation plots for three different cable inserts placed into one channel. 
The signal from
a single detector is split as described in the text and then cable inserts of 1, 2, and 3 feet are
added to the Channel 1 cable run. (b) A test of the timing correction software by using a data
file as in (a). There is no delay through the file, but if we attempt a correction for a shift of the 
peak in either direction, the correlations are spread out in the opposite direction, as expected.}
\end{figure}

We also used this experimental set-up to check that our software computing
the cross-correlation function is working properly. For reducing data from stars, the
changing position of the object across the sky implies that the delay between the two 
channels also changes.  Thus, one must correct for this effect in software, 
even within a single data file
containing a few minutes of data. The lab measurements of course do not have this effect, 
so that if we analyze the lab data as if the peak position is changing, we should observe a smearing 
of the correlation peak in the opposite direction. In Figure 2(b), we show the result when 
assuming the file has a $\pm$1 ns shift in the location of the peak during the exposure time.
(A typical expected shift for our observing sessions would be on the order of 1 to a few ns,
depending on baseline, sky position, and the time spent on the target.)
The (strongly-peaked) original cross-correlation function is then spread into a top-hat function 
in the expected direction and with the correct width.

Lastly, when the time stamps for photons in each channel are to be cross-correlated, a 
coincidence requirement must be established. In general, one would expect that the timing
``bins'' would be chosen so that they have nearly the 
same width as the expected cross-correlation response
discussed just above of 48 ps. However, to verify this, we constructed a simulation program
that outputs data with effective time stamps as short as 1 ps, but with 
the combination of detector timing and electronic jitter
at the measured level. To do this, the cross-correlation function is constructed directly
with 1-ps timing bins as follows. We select a mean number of random correlations per bin and a fraction
of correlated photons for the simulation. The parent probability function for the cross-correlation
 is then a constant 
function with value of the mean number 
of correlations per picosecond plus a Gaussian function with integral determined by the fraction of 
correlated photons
and with width given by the detector plus electronic jitter as discussed above. 
We then select Poisson deviates from the input values in each 1-ps timing bin in the parent distribution.
By rebinning the 1-ps precision cross-correlation to different, coarser bin widths, we can 
investigate the
effect on the SNR. This simulates using
different timing bin widths (coincidence requirements) in the analysis. The results are shown
in Figure 3, where in the left panel, the SNR obtained as a function of timing bin width is
shown for four different simulations. In the right panel, we compare the level of SNR observed
as a fraction of the theoretical value input into the simulation. At the peak value, the SNR
calculated from the data is 89\% of the theoretical value, at a bin width of 64 ps. (It is not 100\% 
of the value one would calculate from Equation 8 because timing jitter sends some correlated 
photons into the wings of the detector response and thus they did not 
appear in the timing bin of the expected peak location.) If a timing bin above 64 ps is used to establish 
coincidence, then more signal counts are included, but at the price of a higher number of 
random correlations. On the other hand, if a bin width less than 64 ps is used, more signal
counts fall outside the criterion for coincidence, also leading to a loss of SNR. For the data
described in this paper, a timing bin of 64 ps will be used in the data reduction for all observations
here.

\begin{figure}[!t]
\figurenum{3}
\plottwo{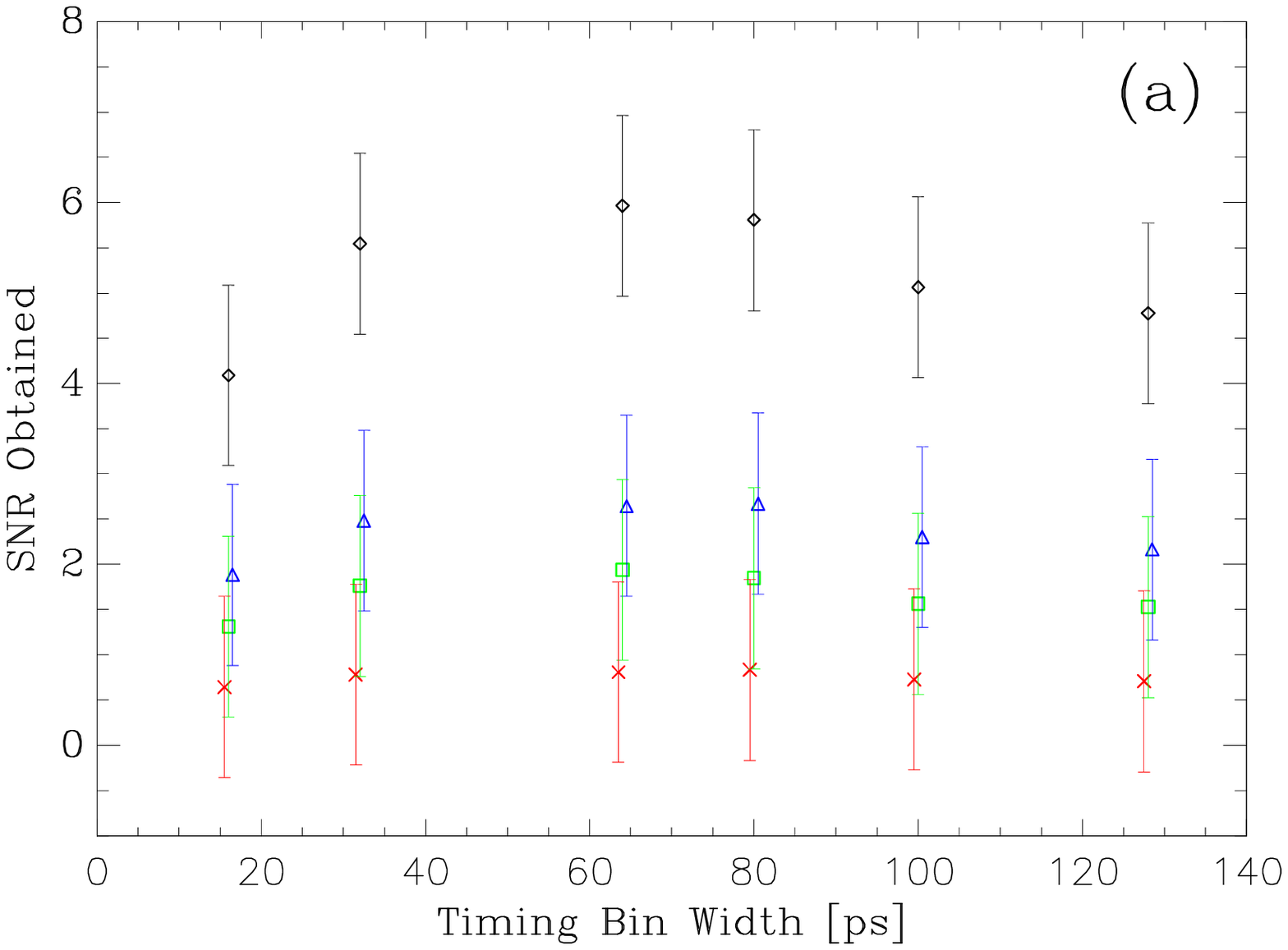}{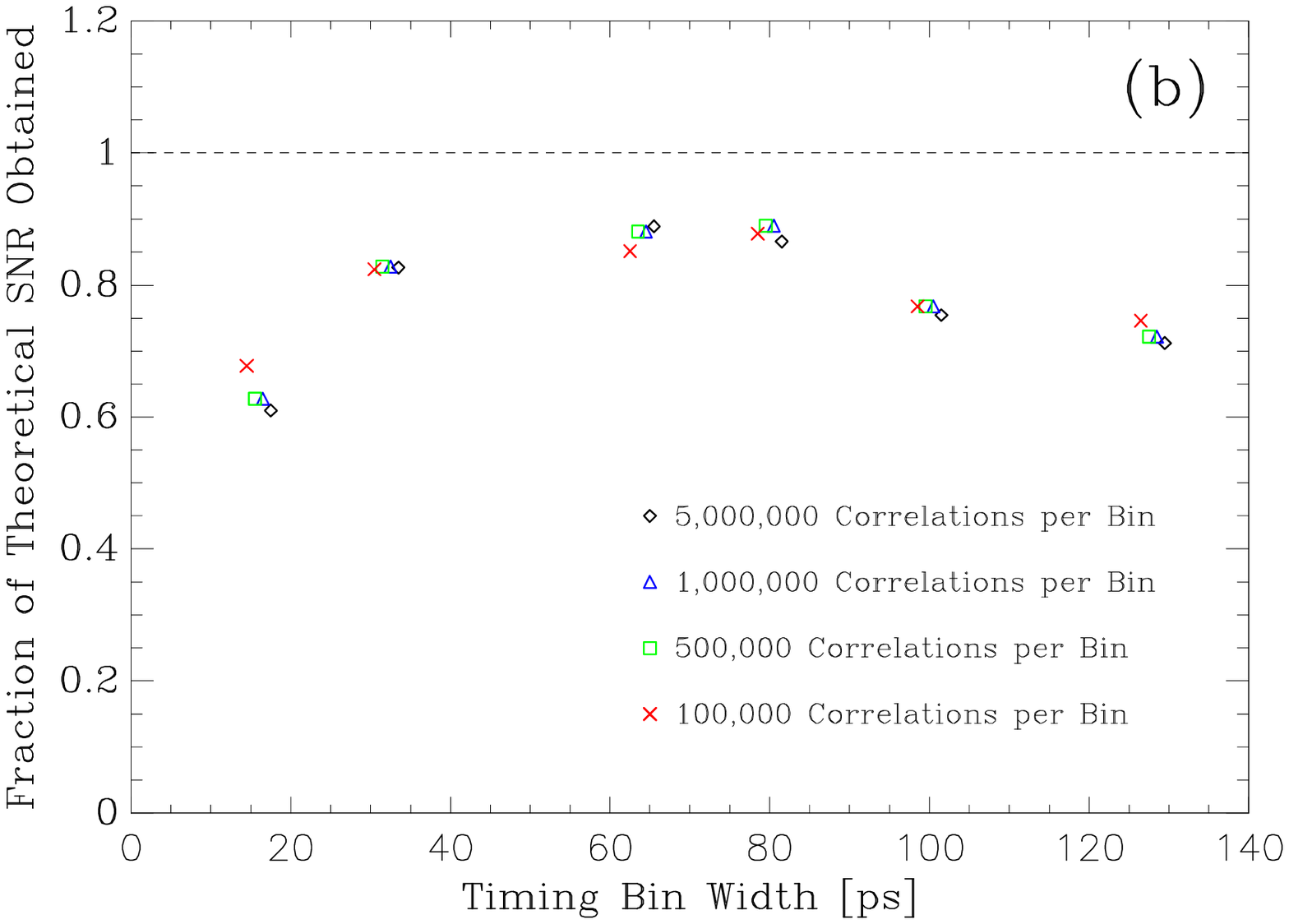}
\figcaption{(a) Signal-to-noise ratio (SNR) obtained in simulation results assuming a timing
uncertainty of 30 ps for each detector and 20 ps uncertainty in the time stamp from the correlator. 
These are shown as a function of the timing bin width used to correlate the data, and span 
a wide range of count rates and total exposure times per night.
(b) The same data shown as a fraction of the theoretical SNR. Here the error bars are removed
for clarity. In both plots, some of the symbols are offset slightly in the
 horizontal direction to avoid overlap.}
\end{figure}

\subsection{Optical Considerations}

Another factor that can reduce the degree of correlations seen is a red leak
or other out-of-band contributions to the light that reaches the detector through the optical system.
By using standard incandescent, fluorescent, and LED light sources in the laboratory, 
we have checked for 
any evidence of measurable out-of-band light reaching the detectors through
the optical harness that includes the filter. Reflecting the light off of a white, flat screen and
covering the detector assembly in black cloth except for the entrance pupil, 
we detect counts through the optical harness, i.e. the detector assembly as it is attached to 
the telescope. For the 633-nm and narrow 532-nm filters, 
count rates obtained in the lab do not show evidence for measurable out-of-band contributions;
the ratio of counts obtained is, after accounting for the difference in spectral power generated
by each lamp at the 
two different wavelengths, largely consistent with the known widths and peak transmissions
of the filters. On the other hand, the wider 532-nm filter has a larger-than-expected count 
rate when viewing the incandescent source, suggesting that this filter has a red leak. This implies
that some percentage of the light we detect from stars in this filter will be uncorrelated, 
even when viewing an unresolved object, decreasing 
the correlated fraction of light by the same percentage as the out-of-band light to the total
amount detected. By using standard (blackbody) emission curves available for
an incandescent source we find we are able to approximately match the observed 
count rate obtained by including a red leak transmission model of perhaps ten percent 
transmission  in the range of 750-1100 nm.

The stellar sources we observe have spectra that peak at much bluer wavelengths
than the standard incandescent laboratory light source. If one assumes a red leak at this
level and computes the out-of-band fraction of light that would be obtained for the spectral 
type of each star we have observed, we find that the out-of-band contribution is at most a 
few percent. The spectra found in the spectral library of \citet{pic98} were used for this calculation.
As a consequence, we will ignore this effect at present. Nonetheless, the wider 532-nm filter
is likely to be useful to us, particularly on the fainter sources we wish to observe, 
so we plan to more fully characterize the transmission at all wavelengths in the future.

An effect that the optical design may have on the width of the correlation peak is to create
a spread due to optical path length differences between paraxial and marginal rays in the
optical system. However, to first order the telescopes themselves will not contribute to this
because the figures of their mirrors are such that they give good image quality at the 
Newtonian focus.
A marginal ray will reflect off of the primary mirror first for a plane wave input
beam, but it travels a longer distance to reach the focal point, largely canceling the 
initial timing difference from the point of reflection.
The optical harness in front of the detector includes two lenses, and one may roughly estimate
the maximum possible 
optical path difference from the telescope focus to the detector focus using a thin lens
approximation. A paraxial ray would travel up to the first lens along the optical axis while a 
marginal ray will travel the hypotenuse of the triangle whose legs are the focal length of the
lens (35 mm) and the radius of the lens (12.5 mm), a difference in distance of 2.2 mm. 
Because the second lens reimages the light onto the detector with the same focal length,
the optical path difference estimate is simply the double of this, namely $\sim$4.4 mm. 
In time, this difference
amounts to a maximum delay of 14 ps in terms of the arrival time of the photons. Averaging over
all rays uniformly distributed in the collimated beam of a circular aperture, the result would
be 7.5 ps with a standard deviation of 4.3 ps. Thus, we will ignore this spread in the work 
presented later in this paper.



\subsection{Geometrical Timing Delay and Telescope Placement}

We define the geometrical timing delay as the timing difference between when
a photon would reach the input pupil on Telescope 1 versus when it would reach Telescope 2.
This delay is therefore equivalent to the path length difference between the two 
telescopes starting from the same wavefront above the telescope apertures, divided by
the speed of light.
If the star being observed is overhead or on the Great Circle that passes through the zenith 
and is perpendicular to the orientation of the baseline, 
then this timing delay would be zero, but depending on the position of the object on the sky,
there can be a substantial difference in arrival times between correlated photons, 
especially given 
our level of timing precision. The 
consequence of this is that, during data taking, the position of the correlation peak in the
cross-correlation will shift as the star's diurnal motion across the sky proceeds.
As discussed by \citet{dyc99}, the timing delay, $\Delta$, may be calculated as

\beq
 \Delta =      - \frac{b_{E}}{c} \cos \delta \sin h
  -\frac{b_{N}}{c} (\sin \phi \cos \delta \cos h - \cos \phi \sin \delta) 
  + \frac{b_{Z}}{c}(\cos \phi \cos \delta \cos h + \sin \phi \sin \delta),
\eeq

\noindent
where ($b_{E}$, $b_{N}$, $b_{Z}$) are the (right-handed) components of the baseline vector ${\bf b}$
in the directions
of East, North, and towards the zenith, $\delta$ is the declination of the star, $h$ is the hour angle
of the observation, and $\phi$ is the latitude of the observing site. 
For other groups performing intensity interferometry observations, the positions of the telescopes
are either fixed or the positions repeatable and well-known. One of the challenges with the 
instrumentation and observing site that
we have is that the telescopes are brought from a storage location
to the observing site for each observing session, and so the placement of the telescopes 
is slightly different from night to night. The above formula allows us to characterize the 
importance of the telescope alignment in the context of our timing precision. In particular,
it is easy to show using Equation 11 that a change of only a few mm in any of the three baseline
components can affect $\Delta$ by 50 ps or more, depending on the sky position of the star.
This is of importance because it is 
comparable to the width of the cross-correlation peak we seek to observe.


Determining the separation between telescopes is a relatively straight-forward procedure
of measuring from a fixed point on one telescope and moving to the same 
point on the other using a retractable metal tape measure. Typically two observers measure
the separation independently and both measures are recorded. By studying the 
differences in separation obtained on  a number of nights, we find that
we have a precision of approximately 1 mm in this parameter for a given placement
of the telescopes. Likewise, the pavement on which we place the telescopes for our
observing has been studied for height differences and these have been characterized.
The height difference is established using a spirit level and placing it in a sequence of adjacent positions along 
the north-south line that we use to align the telescopes. At each position, a digital image of the 
position of the bubble inside the level is recorded. These are then compared with images taken in
the lab of the spirit level when one end is elevated above the other by a known height. In this way,
we can establish the slope of the pavement at each position along the north-south line, and derive the
height of the pavement at each position relative to the starting point. We have also repeated this measurement
three times over a period of approximately two years and obtained the same results each time within 1 mm.
Thus, for a north-south baseline, both $b_{N}$ and $b_{Z}$ can
generally be known at a level where the uncertainty contributes a shift of $\sim$10 ps
or less to the expected location of the cross-correlation peak in the data.

\begin{figure}[!t]
\figurenum{4}
\plottwo{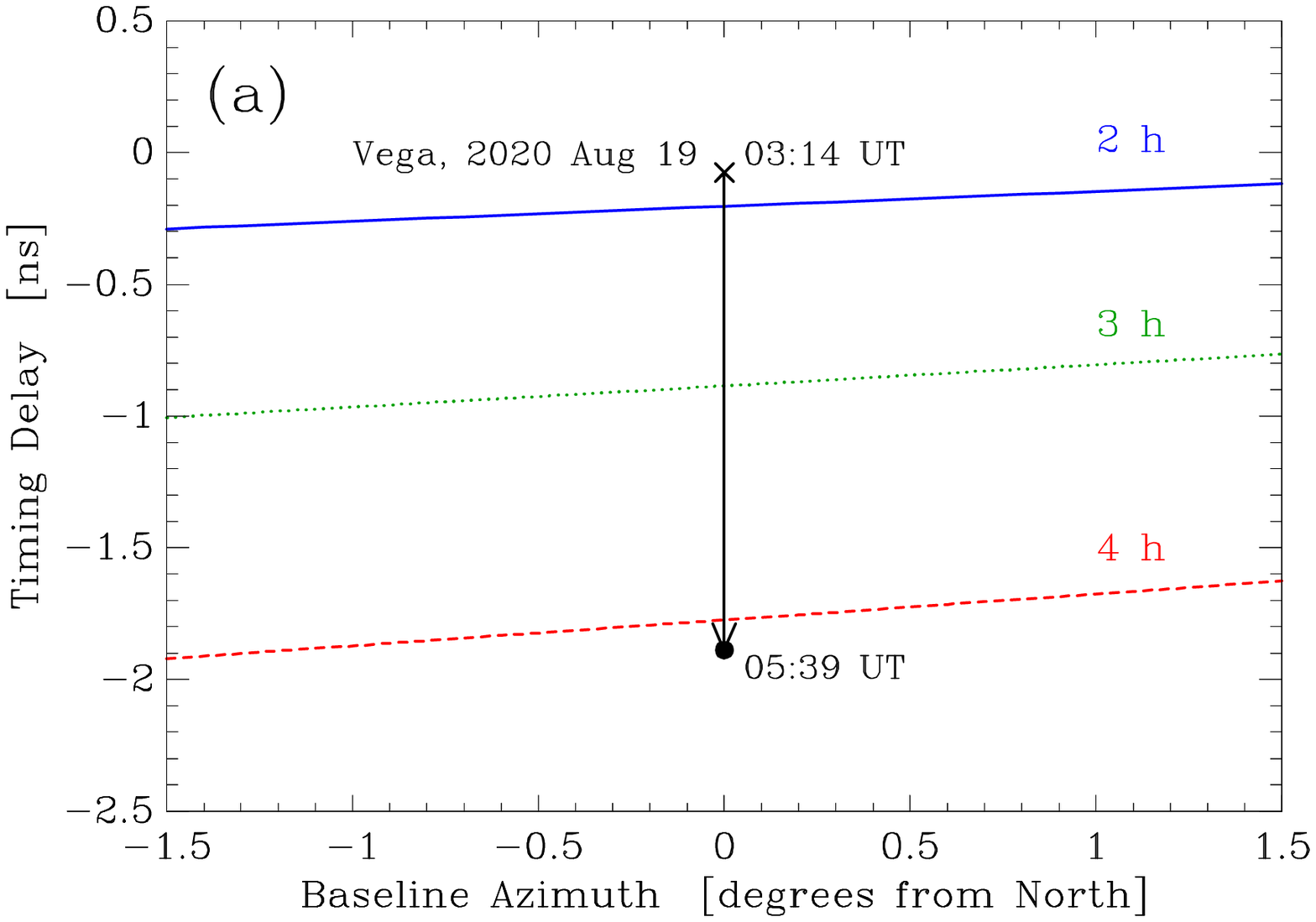}{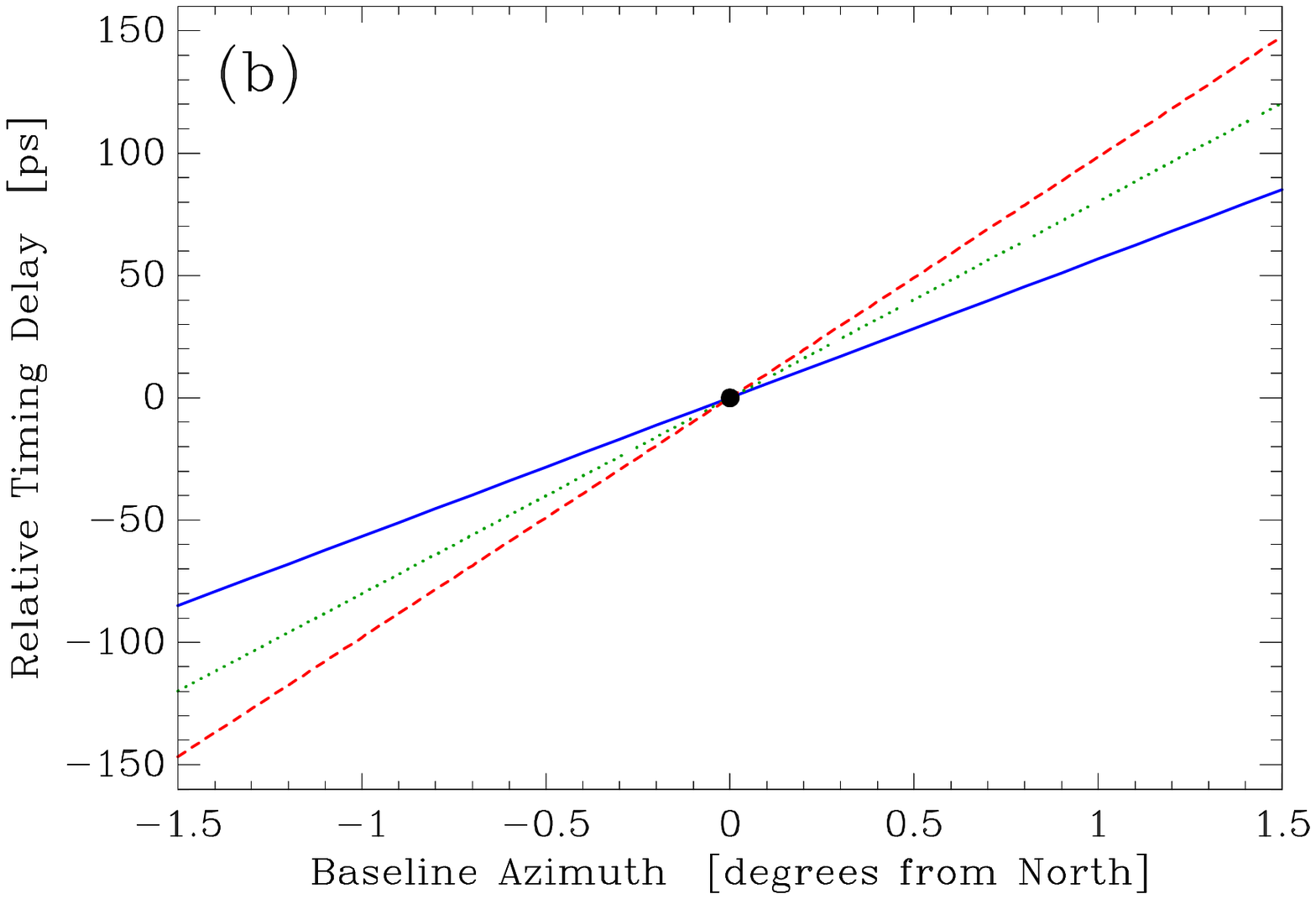}
\figcaption{(a) The timing delay given by Equation 11 in the case of Vega, observed from New Haven,
as a function of the baseline azimuth. Curves are drawn at three different hour angles, 2 h, 3 h, and 4 h,
as indicated by the blue solid curve, the green dotted curve, and red dashed curve respectively. The black 
vertical line indicates the hour angle range during our actual 2020 Aug 19 observation of Vega as an 
example, with the starting 
point indicated with the cross and the end point indicated with the filled circle.
(b) The relative timing delay, if a north-south baseline is assumed and the timing corrections indicated
along the black vertical line are applied for the entire observing sequence, as a function of baseline azimuth. 
Curves for the same three hour angles are shown,
where again the blue solid curve represents an hour angle of 2 h, the green dotted curve represents
3 h, and the red dashed curve represents 4 h.}
\end{figure}

However, given our observing site, measuring the angle made with north by the telescope 
baseline each night is a more difficult task. We have identified the direction to north on the pavement
using a plumb bob and marking its shadow at solar noon, but the painted markings are
about 2 cm wide. When the telescopes are placed, we must sight down along the telescope
base to determine the alignment with the paint mark. We estimate that this can
typically be done only to within several mm at present. In this arrangement, we therefore
have a much larger uncertainty in $b_{E}$ than in the other two components of the baseline.

\begin{figure}[!t]
\figurenum{5}
\plottwo{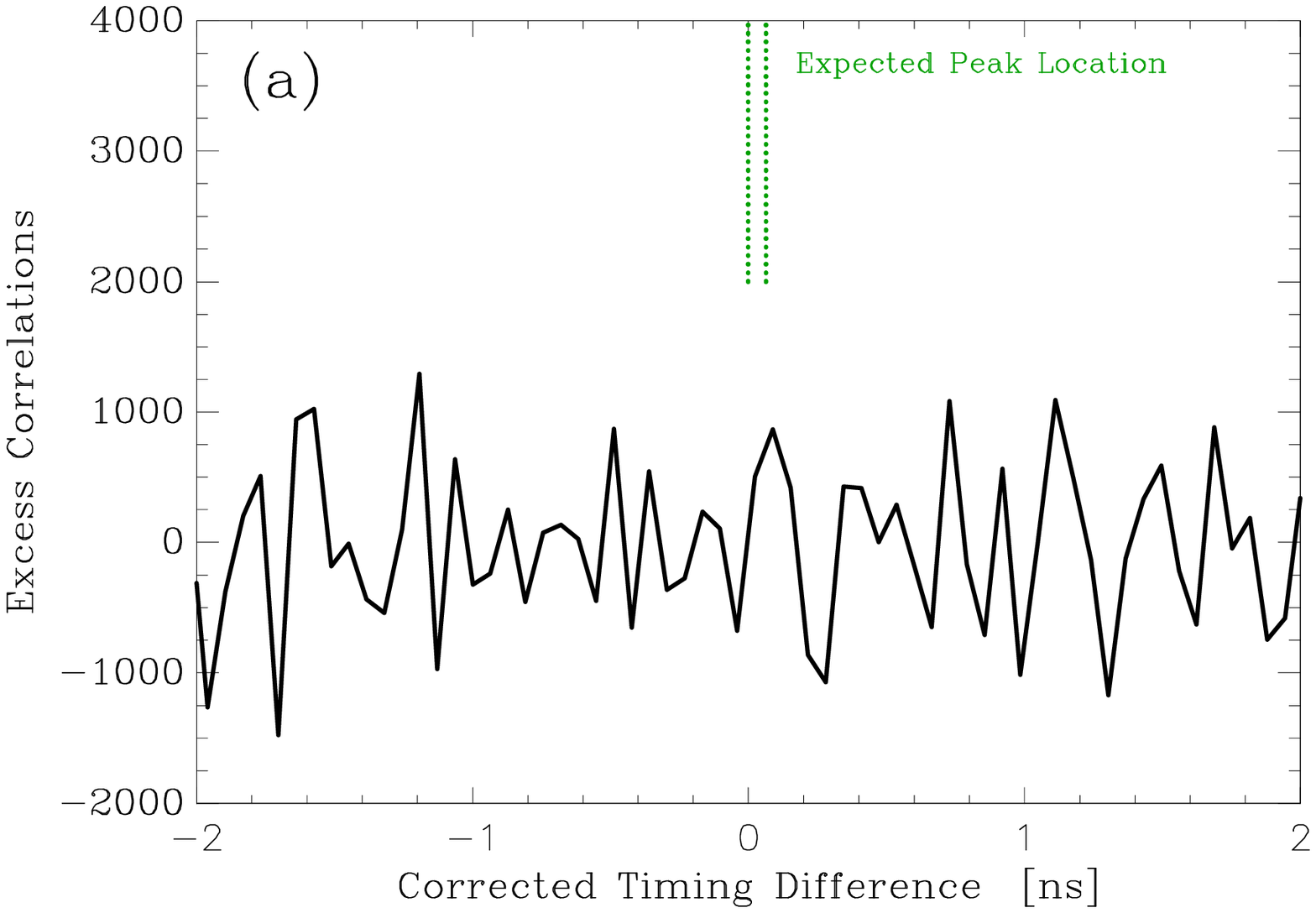}{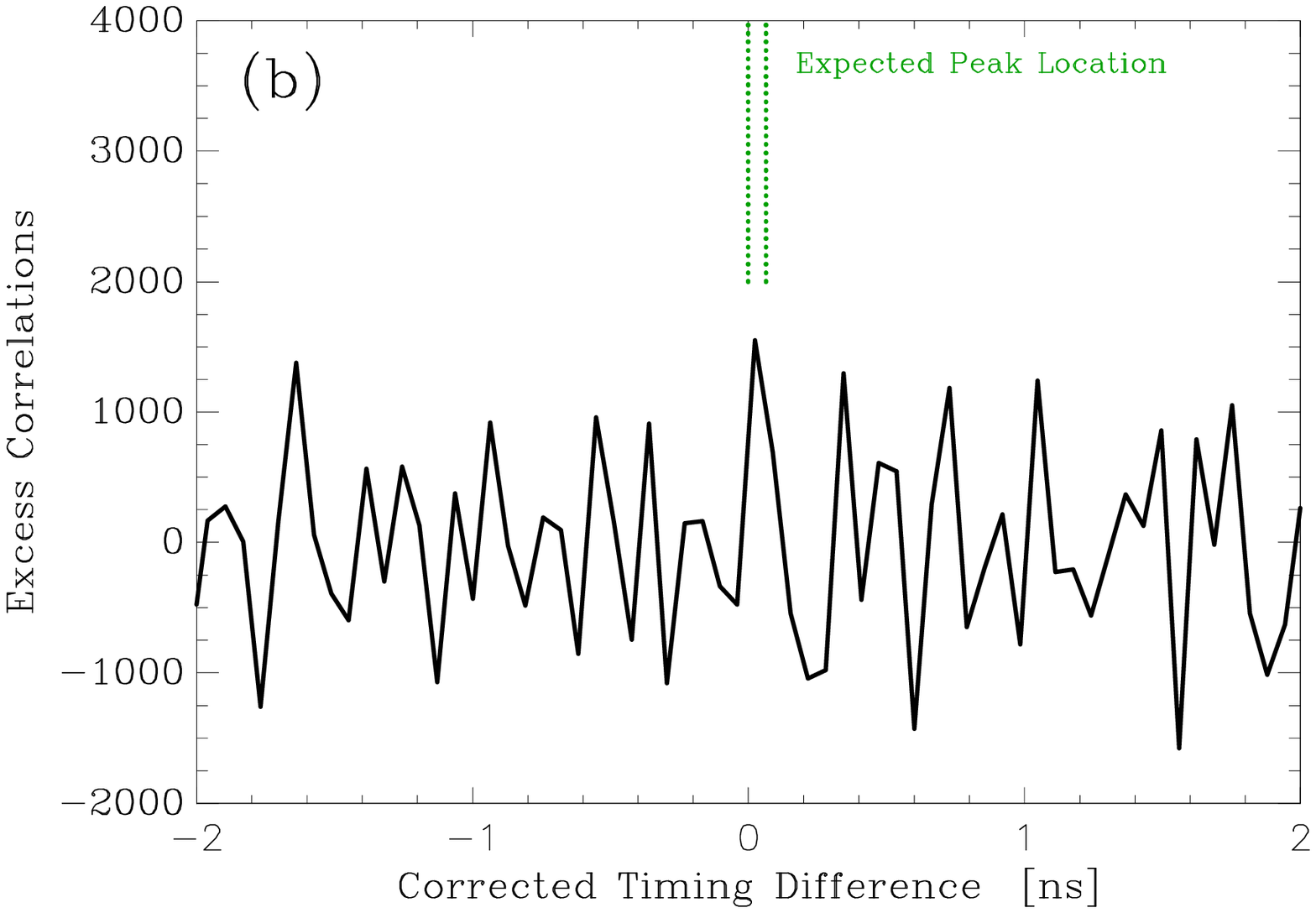}

\vspace{0.2cm}
\plottwo{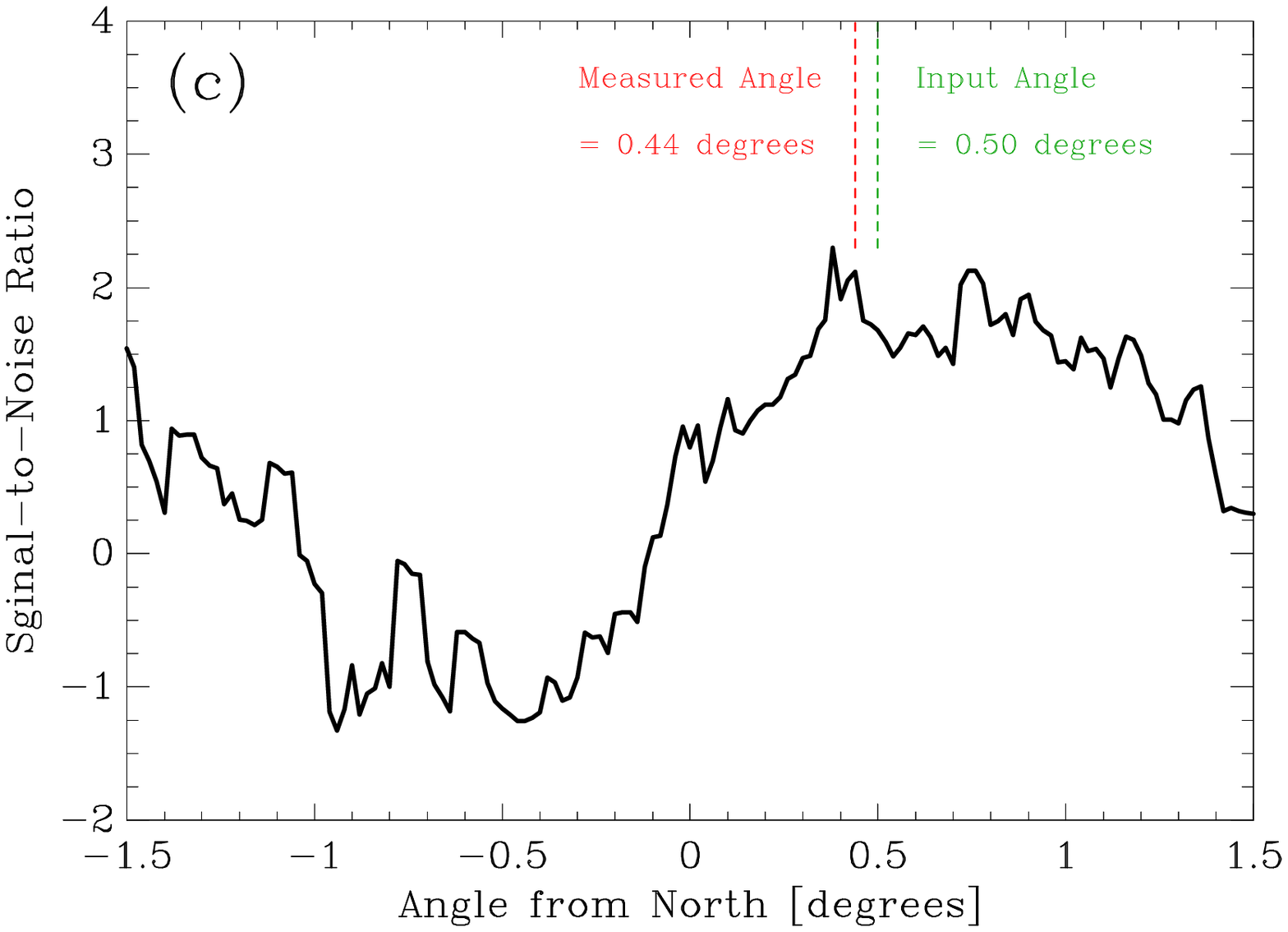}{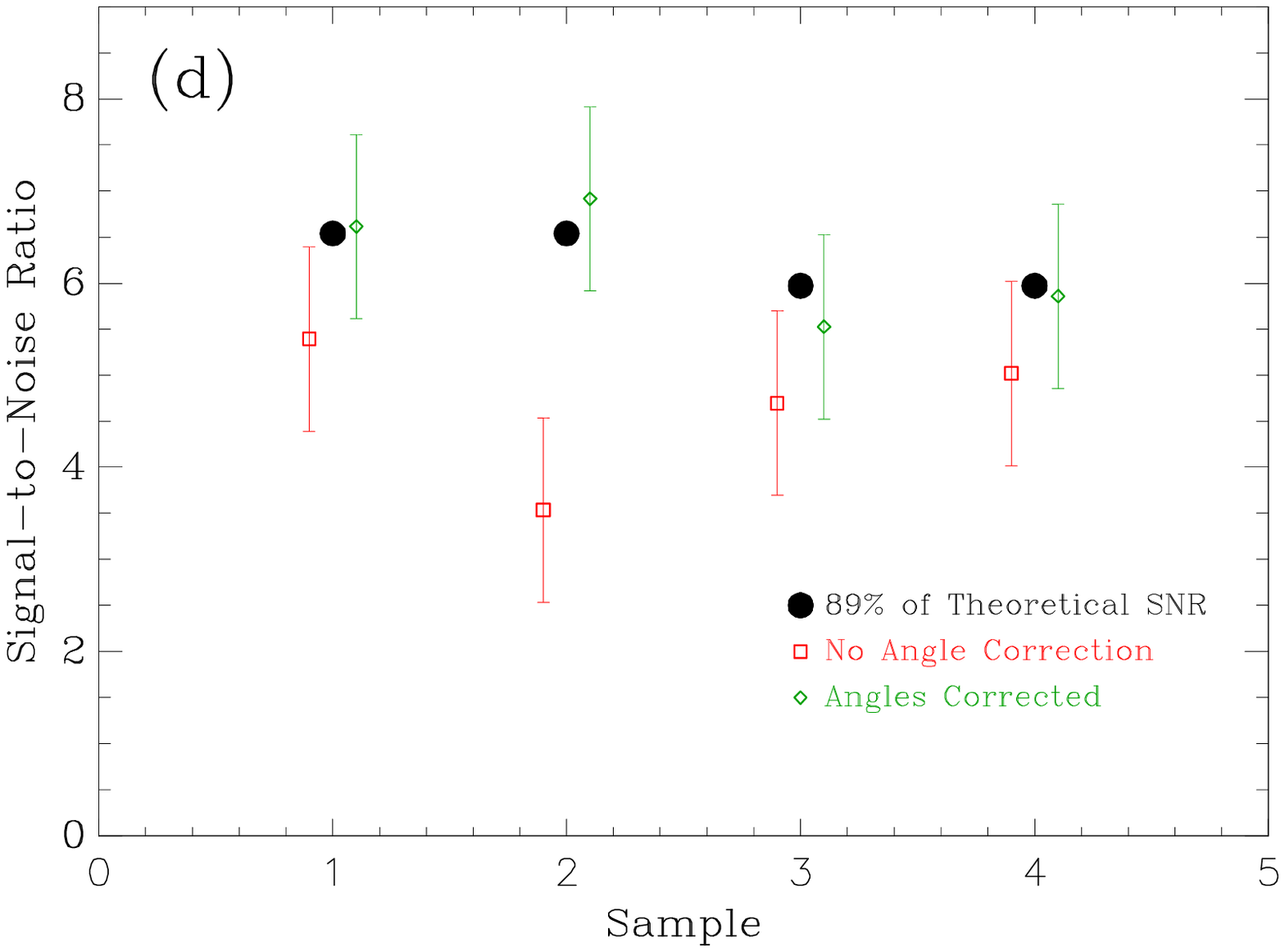}
\figcaption{
(a) Simulation using the hour angle variation of Vega for 2020 Aug 19 and computing
the final, corrected cross-correlation plot assuming a north-south baseline. The 
SNR here in the expected peak location is 0.80, and the input baseline azimuth was 0.5 degrees. 
(b) The same data, but with timing corrections made assuming a baseline azimuth of
0.44 degrees. Here the SNR is 2.12. 
(c) The SNR obtained for the simulation as a function of baseline azimuth. 
(d) Total SNRs obtained for the samples created from the summation of 
ten simulations in each case. Black circles indicate 89\% of the theoretical correlation 
(i.e. what we would expect to
recover using a timing bin of 64 ps given the result in Figure 3), 
the red squares indicate the results of the summations if
the input angles are not estimated and corrected, and the green diamonds indicate the result of
measuring the input angle as in (c) and then co-adding. This process returns the SNR to the level
expected.}
\end{figure}

To study this effect further, 
we first plot the expected timing difference for a typical observing sequence as a function
of the baseline azimuth in Figure 4(a). We use the declination of Vega, latitude of New Haven, and
we show the timing delay ($\Delta$ from Equation 11) obtained for three different hour angles. (This range in 
hour angle
is comparable to the observing sequence of Vega on 2020 August 19 that will be discussed in the 
next section, and that observing sequence is illustrated in the plot.) In Figure 4(b), we show
the relative timing difference, if the timing delay were to be corrected based on the assumption of a north-south
baseline. At a $0^{\circ}$ baseline azimuth, the relative delay is always zero as the timing shifts used throughout
the sequence are always correct, but if the baseline azimuth differs from zero, we see that there is a
shift in the final location of the peak, which is more severe at higher hour angles. The 
effect of this in the final cross-correlation function computed from the timing data of the entire observing
sequence is twofold: first, 
there is an average shift of the cross-correlation peak away from the expected location, and second,
the peak is spread out because of the variation in the shift as the hour angle changes.

Next,
we constructed a simulation of the data taking process that inputs 
a baseline azimuth offset and outputs a collection of 25 individual cross-correlations, assumed to be taken
over two hours, building in
a timing shift throughout that is based on 
the path of Vega across the sky during that time.
We then compute the cross-correlation plot for the entire simulated data set,
correcting for $\Delta$.
In Figures 5(a) and (b), we show two examples of final cross-correlations
obtained in this way, for an input baseline azimuth of $0.5^{\circ}$ and a count rate of 1 MHz. 
For (a), the timing delays used to correct
the data assume a baseline azimuth of $0^{\circ}$, while in (b), the shifts assume a baseline azimuth of $0.44^{\circ}$.
We see that, since the first case uses shifts based on an azimuth that is $0.5^{\circ}$ away from the true input angle, the peak appears
shifted and spread out. On the other hand, when a value close to the actual azimuth is used to determine
the timing corrections, the peak is narrower and appears at the predicted location. 
We can then vary the baseline azimuth through a range of $-1.5^{\circ}$ to $+1.5^{\circ}$,
recomputing the final
cross-correlation in each case. Figure 5(c) plots the SNR of these final cross-correlation peaks as a function
of baseline azimuth.
The original input azimuth is marked 
(in green), and it can be seen that a broad
peak in SNR occurs in this vicinity as  more correlation counts are placed at the expected peak location
when applying timing shifts generated from those azimuth values. 
We constructed a routine to estimate the location of this
peak from a smoothed version of the SNR curve, and the location of that is also shown in 
the graph (in red). While typical results show that the input azimuth and the recovered value can differ 
by up to 0.1-0.2$^{\circ}$ given the expected SNRs per night, the ``tuning" of 
the azimuth value in software can nonteheless help recover SNR that is lost
if the assumption of a 0$^{\circ}$ azimuth is made. In panel (d) of the figure, we show the results
of four samples of simulations (simulation families) done at two different count rates, 
where measuring the azimuth from the data
results in a SNR that is consistent with expectations, once the data is binned in 64-ps 
increments, whereas if a north-south baseline is assumed, some signal-to-noise is lost.



\section{Observations}

In this paper, we discuss observations obtained on 16 nights on campus at SCSU and 3 nights
where the timing correlator and detectors were taken to Anderson Mesa in Arizona and used with
the 1.0-m Hall Telescope and 1.8-m Perkins Telescope at Lowell Observatory.

\begin{deluxetable}{llrrccclc}[!t]
\tablewidth{0pt}
\tablenum{2}
\tablecaption{SCSU Observing Summary}
\tablehead{
\colhead{Object} & Bayer & HR & HD &
\colhead{Date} &
\colhead{Tel. Sep.\tablenotemark{a}} &
\colhead{Filter} & \colhead{Northern} & \colhead{Total Observing}\\
&Desig. &&& (UT) & (m) & ($\lambda_{0}$/$\Delta \lambda$, nm) & \colhead{Configuration} &Time (hr)
}
\startdata
Altair & $\alpha$ Aql & 7557 & 187642 & 2016 Aug 09 & 3.01 & 532/3 & Tel 2, +1ft, Ch 1 & 0.61 \\
Arcturus & $\alpha$ Boo & 5340 & 124897 & 2017 Jun 04 & 2.099 & 532/3 & Tel 2, Ch 1 &  0.34 \\
Arcturus & $\alpha$ Boo & 5340 &124897 & 2021 Jun 07 & 2.497 & 532/1.2 & Tel 2, Ch 1 & 1.10  \\
Arcturus & $\alpha$ Boo & 5340 &124897 & 2021 Jun 18 & 2.488 & 532/1.2 & Tel 2, Ch 7\tablenotemark{b} & 1.33  \\
Arcturus & $\alpha$ Boo & 5340 &124897 & 2021 Aug 07 & 2.488 & 532/1.2 & Tel 2, Ch 7\tablenotemark{b} & 0.58  \\
Deneb & $\alpha$ Cyg & 7924 & 197345 & 2020 Aug 19 & 2.513 & 532/3 & Tel 2, Ch 0 & 0.58 \\
Deneb & $\alpha$ Cyg & 7924 & 197345 & 2020 Aug 25 & 2.497 & 532/3 & Tel 2, Ch 0 & 1.24 \\
Deneb & $\alpha$ Cyg & 7924 & 197345 & 2020 Aug 26 & 2.493 & 532/3 & Tel 2, Ch 0 & 1.54 \\
Deneb & $\alpha$ Cyg & 7924 & 197345 & 2021 Jun 07 & 2.497 & 532/1.2 & Tel 2, +1ft, Ch 1 & 0.67 \\
Polaris & $\alpha$ UMi &  424 & 8890 & 2020 Sep 12 & 9.167 & 532/3 & Tel 2, Ch 0 & 2.17 \\
Vega & $\alpha$ Lyr & 7001 & 172167 & 2016 Jun 23 & 3.05 & 532/3 & Tel 1, +1ft, Ch 1 & 0.50 \\
Vega & $\alpha$ Lyr & 7001 & 172167 & 2016 Jul 27 & 2.31 & 532/3 & Tel 2, +1ft, Ch 1 & 0.51 \\
Vega & $\alpha$ Lyr & 7001 & 172167 & 2016 Sep 14 & 2.98 & 532/3 & Tel 2, $-$1ft, Ch 0 & 0.73 \\
Vega & $\alpha$ Lyr & 7001 & 172167 & 2017 Jun 04 & 2.099 & 532/3 & Tel 2, +1ft, Ch 1\tablenotemark{d} & 2.91 \\
Vega & $\alpha$ Lyr & 7001 & 172167 & 2017 Jun 21 & 2.331 & 532/3 & Tel 2, Ch 1 & 1.93 \\
Vega & $\alpha$ Lyr & 7001 & 172167 & 2017 Aug 09 & 2.532 & 532/3 & Tel 2, Ch 1 & 1.39 \\
Vega & $\alpha$ Lyr & 7001 & 172167 & 2018 May 30 & 2.502 & 532/3 & Tel 2, +2ft, Ch 7\tablenotemark{b}  & 0.65 \\
Vega & $\alpha$ Lyr & 7001 & 172167 & 2020 Aug 19 & 2.513 & 532/3 & Tel 2, +1ft, Ch 1\tablenotemark{c} & 2.08 \\
Vega & $\alpha$ Lyr & 7001 & 172167 & 2020 Aug 25 & 2.497 & 532/3 & Tel 2, +1ft, Ch 1 & 1.24 \\
Vega & $\alpha$ Lyr & 7001 & 172167 & 2020 Aug 26 & 2.493 & 532/3 & Tel 2, +1ft, Ch 1 & 1.82 \\
Vega & $\alpha$ Lyr & 7001 & 172167 & 2020 Sep 24 & 2.496 & 633/1 & Tel 2, +1ft, Ch 1 & 3.70 \\
Vega & $\alpha$ Lyr & 7001 & 172167 & 2021 Jun 07 & 2.497 & 532/1.2 & Tel 2, +1ft, Ch 1 & 2.09 \\
Vega & $\alpha$ Lyr & 7001 & 172167 & 2021 Jun 18 & 2.488 & 532/1.2 & Tel 2, Ch 7\tablenotemark{b} & 1.83 \\
Vega & $\alpha$ Lyr & 7001 & 172167 & 2021 Aug 07 & 2.488 & 532/1.2 & Tel 2, Ch 7\tablenotemark{b} & 3.00 \\
\enddata
\tablenotetext{a}{All observations taken with a N-S orientation between the telescopes.}
\tablenotetext{b}{The 8-channel HydraHarp timing module was used on this night.}
\tablenotetext{c}{Files 0-3 had the 1-ft cable on Channel 0 (Tel 1).}
\tablenotetext{d}{Files 0-5 had no 1-ft cable on Channel 1 (Tel 2).}
\end{deluxetable}

\subsection{On-Campus Observations}

The observations taken on-campus have been obtained over a period of nearly 5 years, 
with the observing site being a relatively secluded area of blacktop near one of our
science buildings. 
A listing of the 
observations at SCSU is given in Table 2. 
We have so far only observed with a north-south arrangement
of the telescopes, and to place these telescopes, we align them with paint markings
that were made for that purpose as described in the previous section. 
Data were taken on several other nights that
are not in Table 2, but on these occasions, either we did not collect enough data to 
warrant inclusion, or on two occasions, the electronics exhibited noise that was uncharacteristic
of the system and which led to strong systematic signatures in the cross-correlation function.

A typical observing sequence consists of placing the telescopes and making the baseline
measurements needed for the reductions, checking the collimation of each telescope
with a laser collimator, and then sighting on two stars with an eyepiece so that the 
control units on each telescope 
can begin tracking the diurnal motion. With the telescopes
tracking, we move to the star to be observed, and check the placement of the star
in the eyepiece to make sure it is coincident with the center of the finder. 
We then replace the eyepiece with the SPAD detector and optical 
harness, and attempt to place the star image on the detector by guiding the telescopes
manually. This is done simply by trying to maximize the count rate. Positions of good focus 
were determined early on in our work with the system, and so the detectors are set at
the location of good focus as we mount them, but to further maximize count rates once on
the target, we would slightly adjust the focus as needed. Generally, we begin data taking 
when a count rate of 1 MHz or higher was seen on each channel, and collect data in 5-minute 
intervals. Because our telescopes have no autoguiders at this stage, it is necessary to 
manually guide using the telescope hand paddle throughout the observation sequence.
As with many alt-az telescopes, our telescopes track poorly near the zenith. Typically,
at an altitude of 85$^{\circ}$ or higher, manual guiding becomes difficult, and one must
wait for the star to pass through the meridian to continue observing.

We first discuss the count rates achieved for the on-campus set-up. Figure 6 shows two
typical examples of the instantaneous count rates detected, the first for Vega and the 
second for Polaris. Both observations use the wider 532-nm filter. It is seen here that, 
although Vega is brighter than Polaris by approximately 2 magnitudes, the average count
rates for the two observations are comparable. This is due to the fact that the SPAD detectors
and the timing correlator have relatively low maximum count rates that can be read out. 
For the detector, the cause of this is the relatively long deadtime. However, when reading out
in time tag mode, the timing correlator also has a limitation in how fast it can write events. 
Events are stored in a first-in-first-out (FIFO) buffer as they are transferred to the host computer, 
and if a large burst of events occurs
in a short amount of time, then the FIFO buffer can be overfilled. This causes data taking to
stop. There is a good chance of a FIFO overrun in a file if the average count rate 
exceeds approximately
2 MHz, and so, even on bright objects, the count rate was kept below this value to prevent 
having to restart files often. (This was done by defocusing and/or moving the star slightly 
off the center position relative to the detector active area.) In the most recent data taken 
(in 2021), we have been able to upgrade
our host computer, and it handles data faster. Thus, moving forward, we expect to be able to 
increase our average count rates accordingly. However, at present, 
we see that count rates in excess of 1 MHz can be obtained on 2nd magnitude stars
with SCSI.

\begin{figure}[!t]
\figurenum{6}
\plottwo{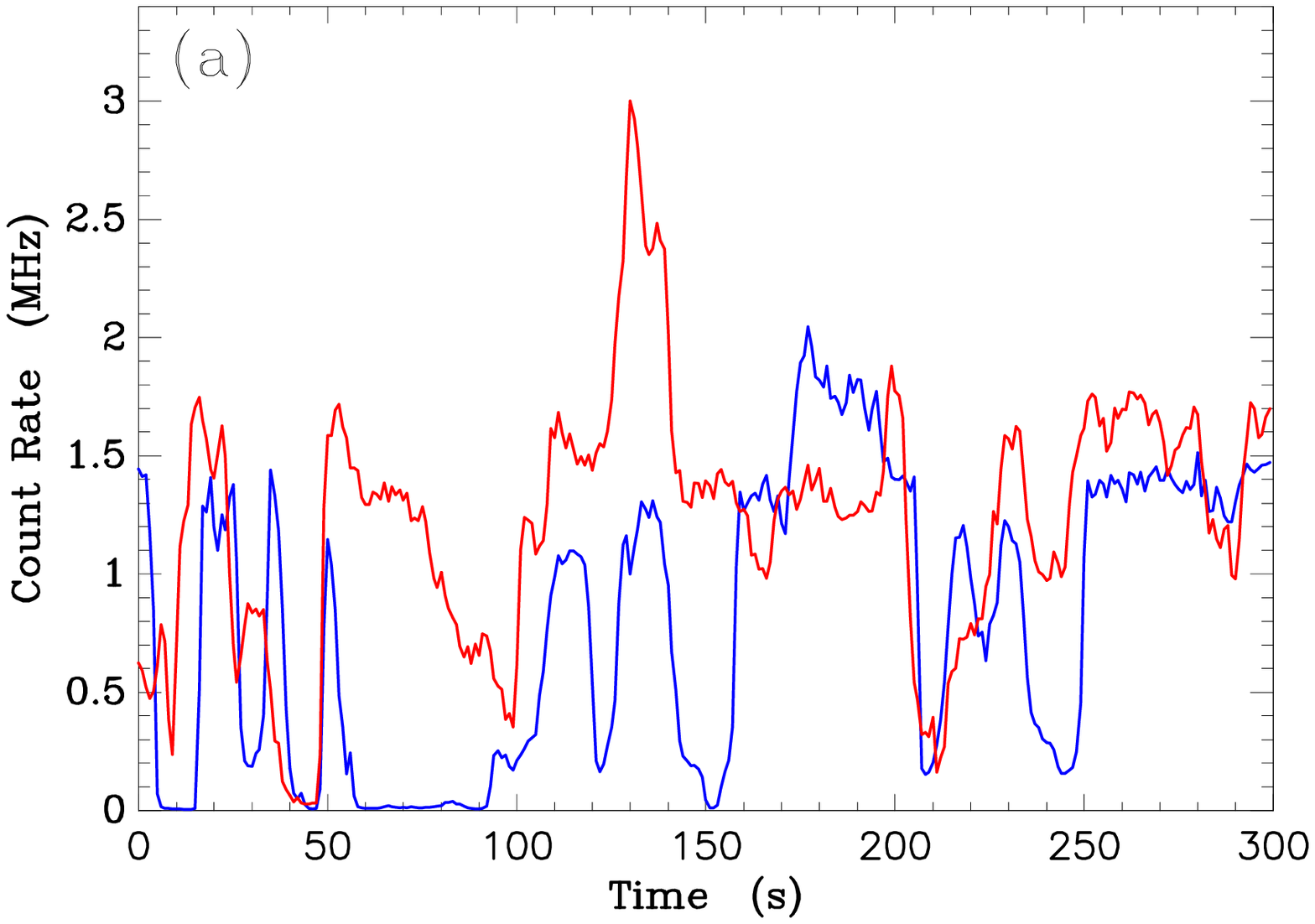}{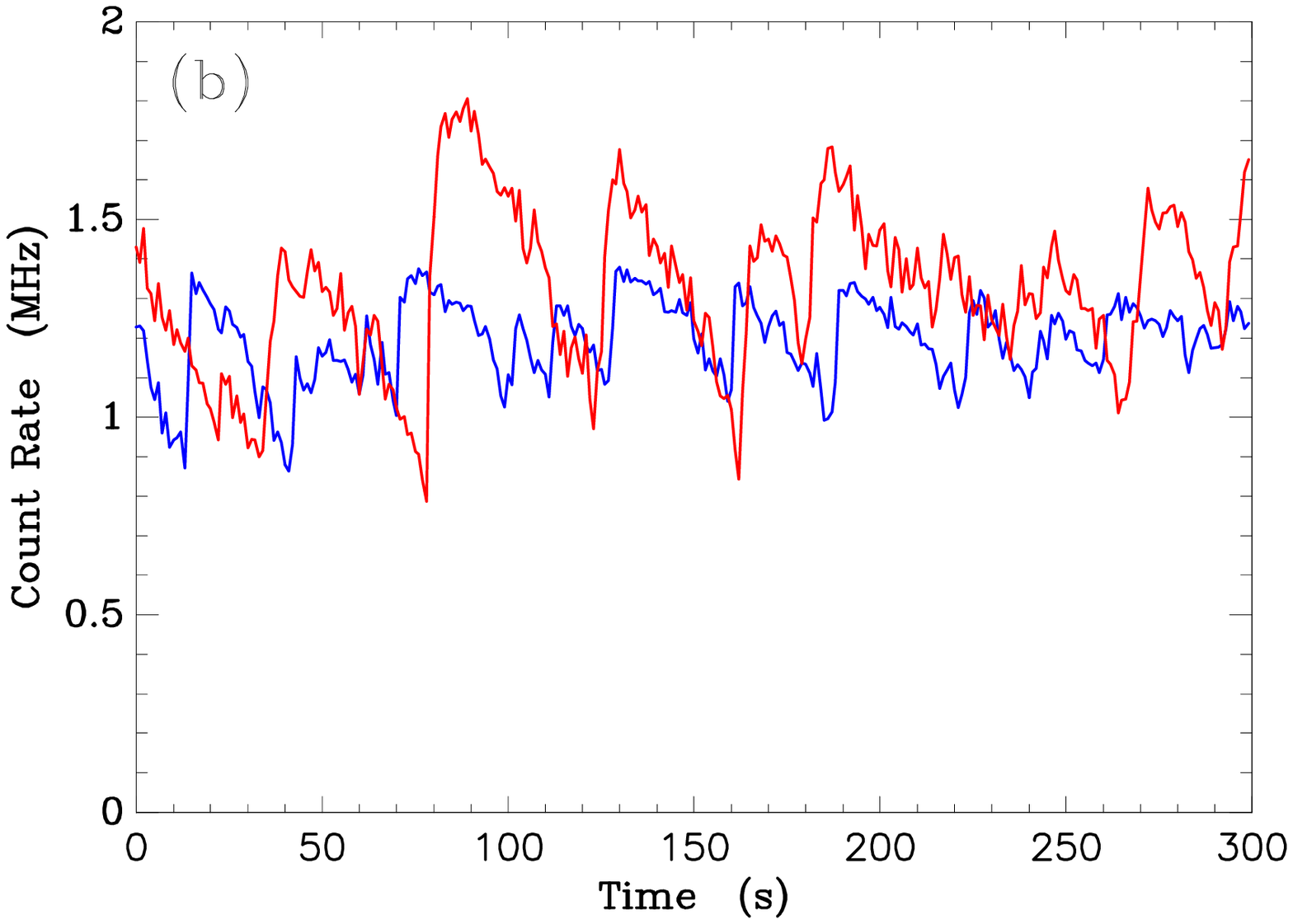}
\figcaption{(a) Instantaneous count rates observed for two stars using the wider
532-nm filter. (a) Vega, on 2020 Aug 19, and (b) Polaris observed on 2020 Sep 12. Blue
curves represent the count rate in Channel 0 of the correlator, and red curves show
the result for Channel 1.}
\end{figure}

\begin{figure}[!t]
\figurenum{7}
\plottwo{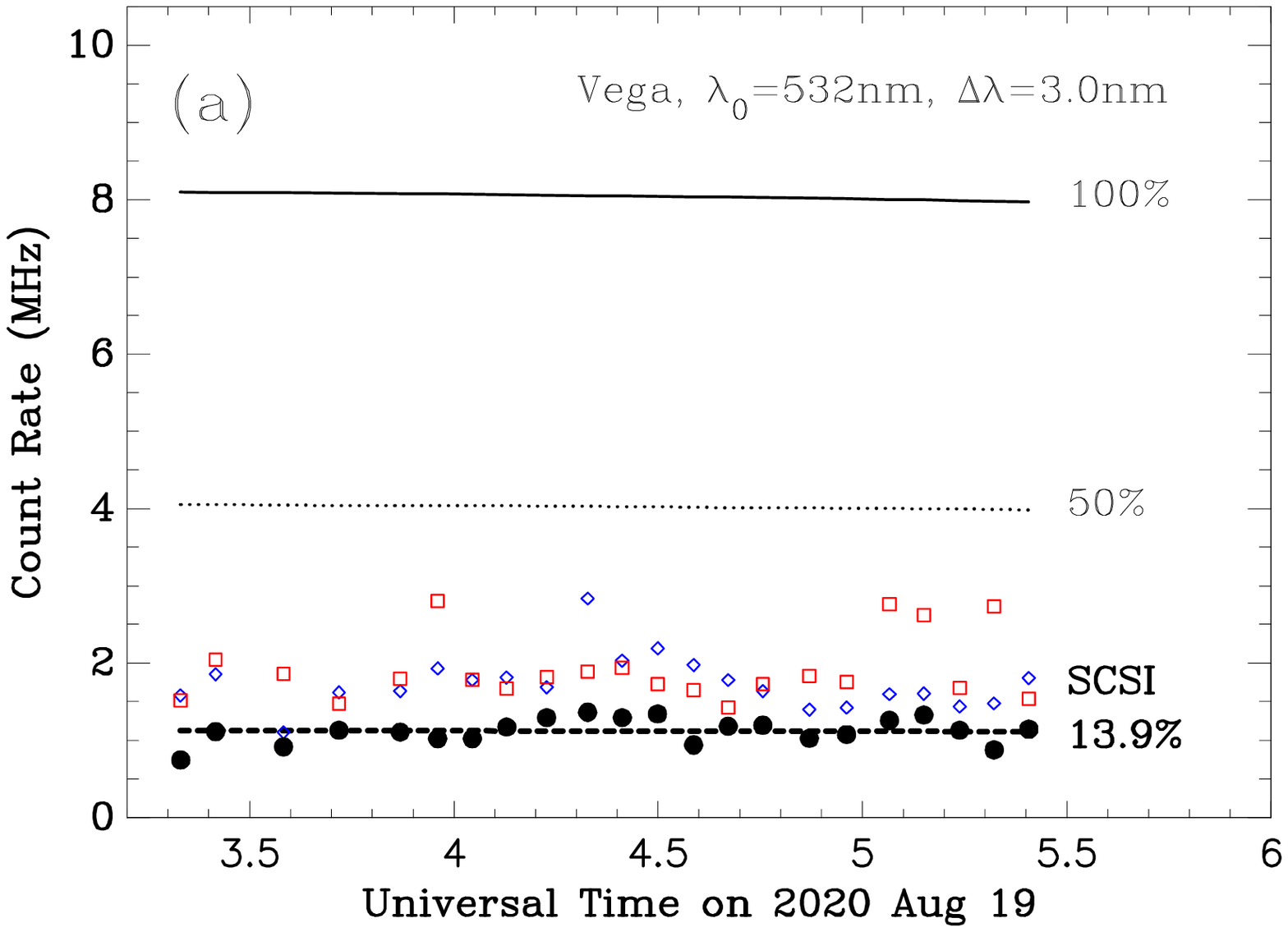}{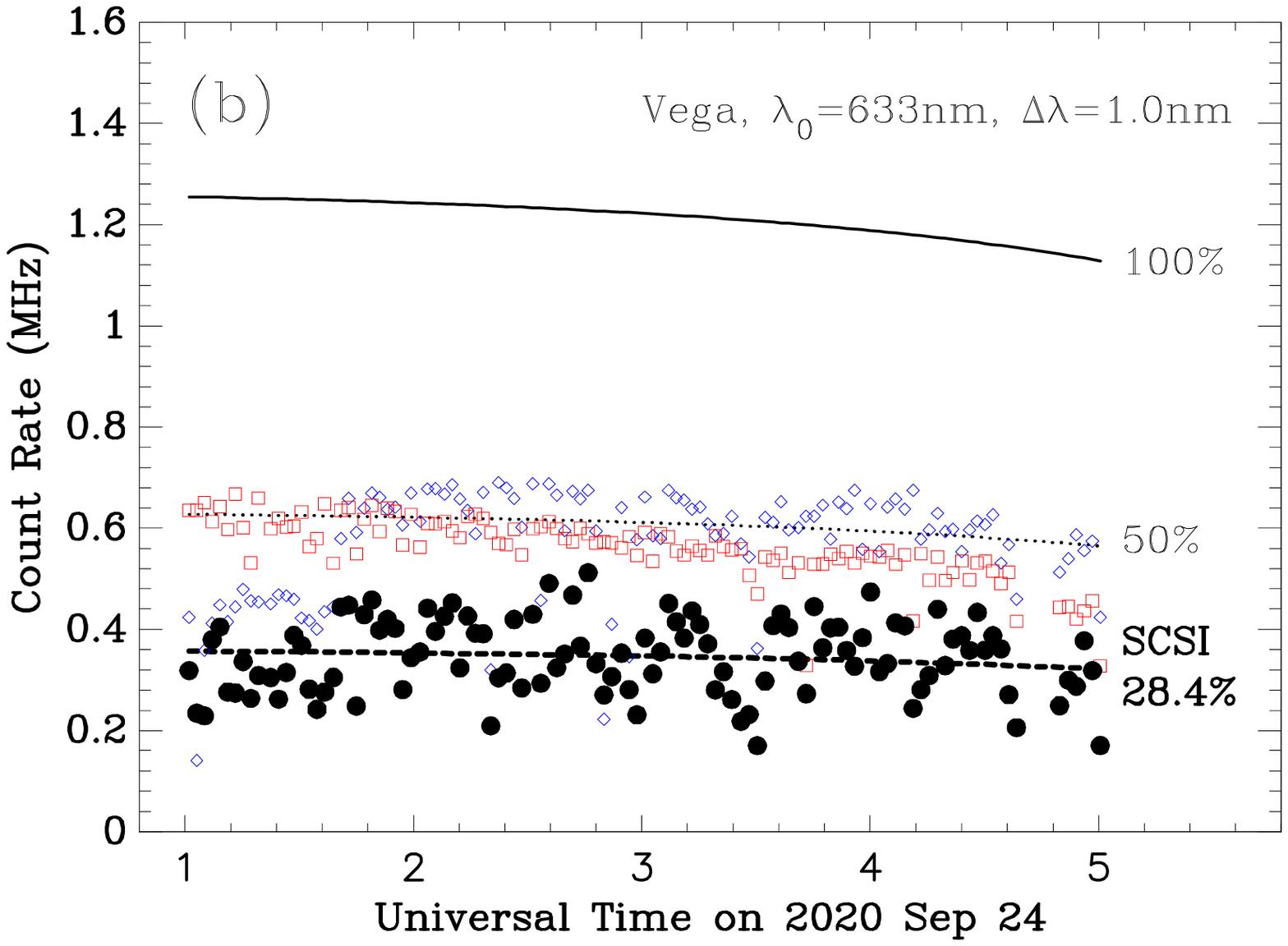}

\vspace{0.4cm}
\plottwo{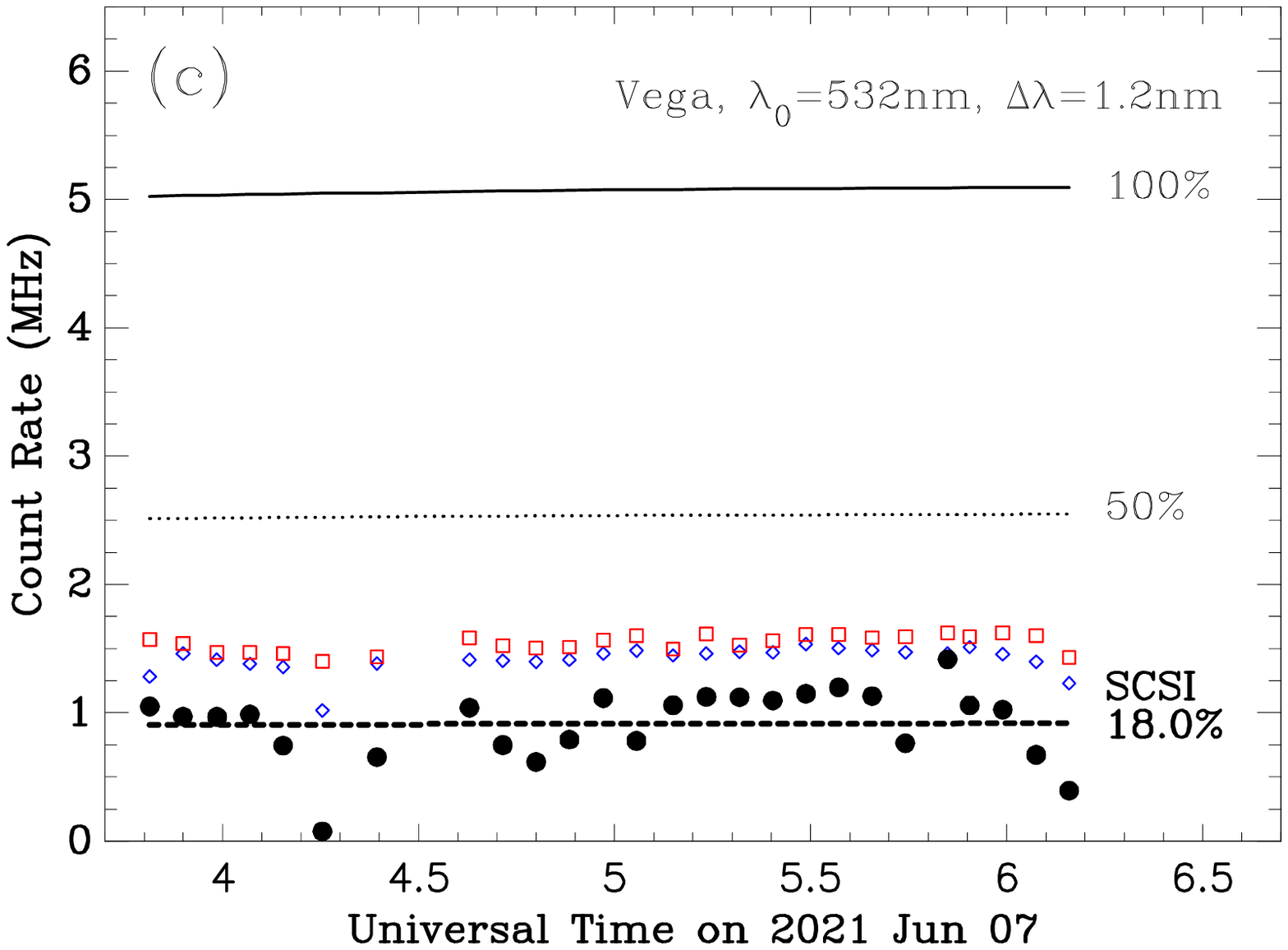}{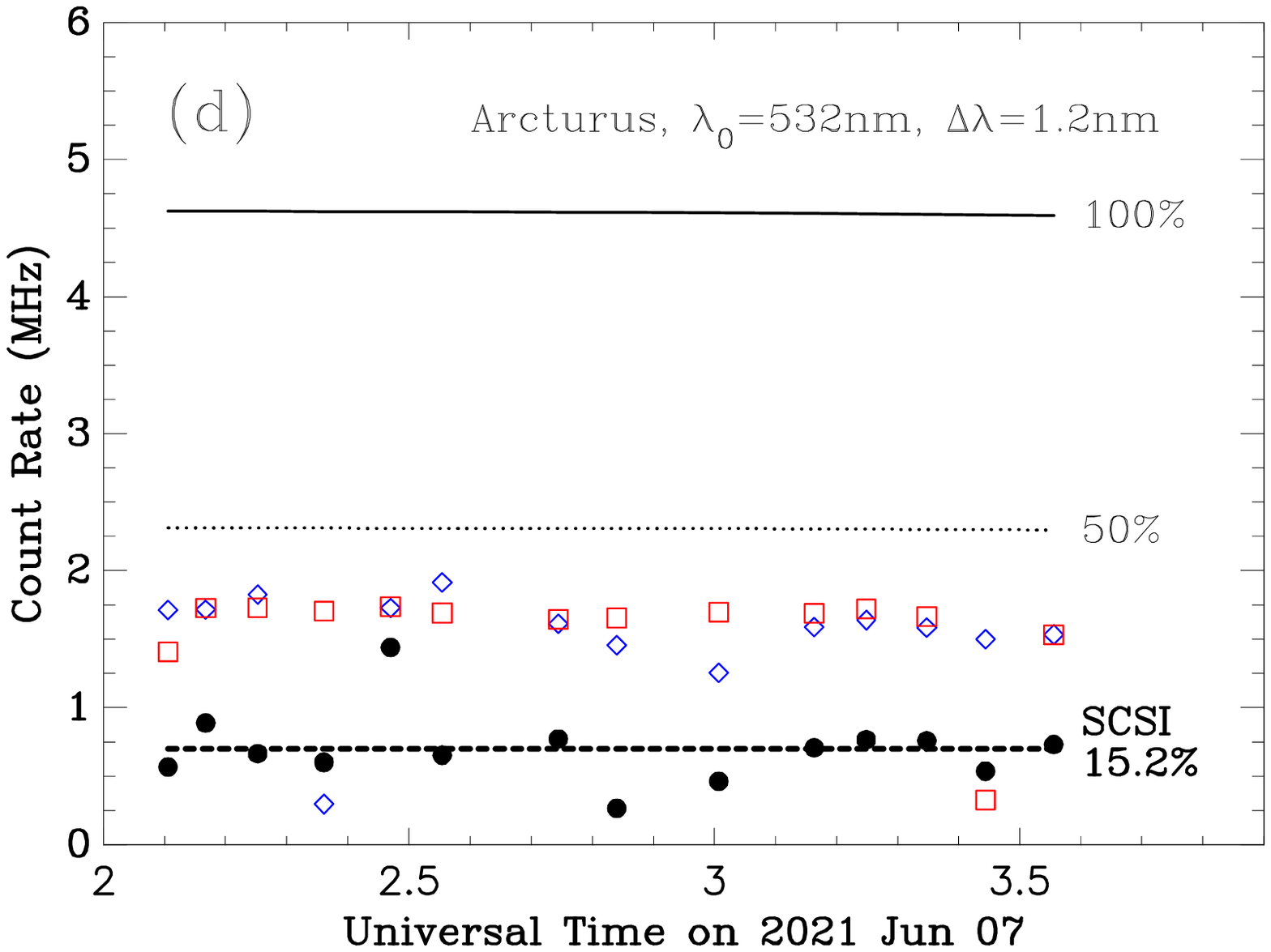}

\vspace{0.4cm}
\plottwo{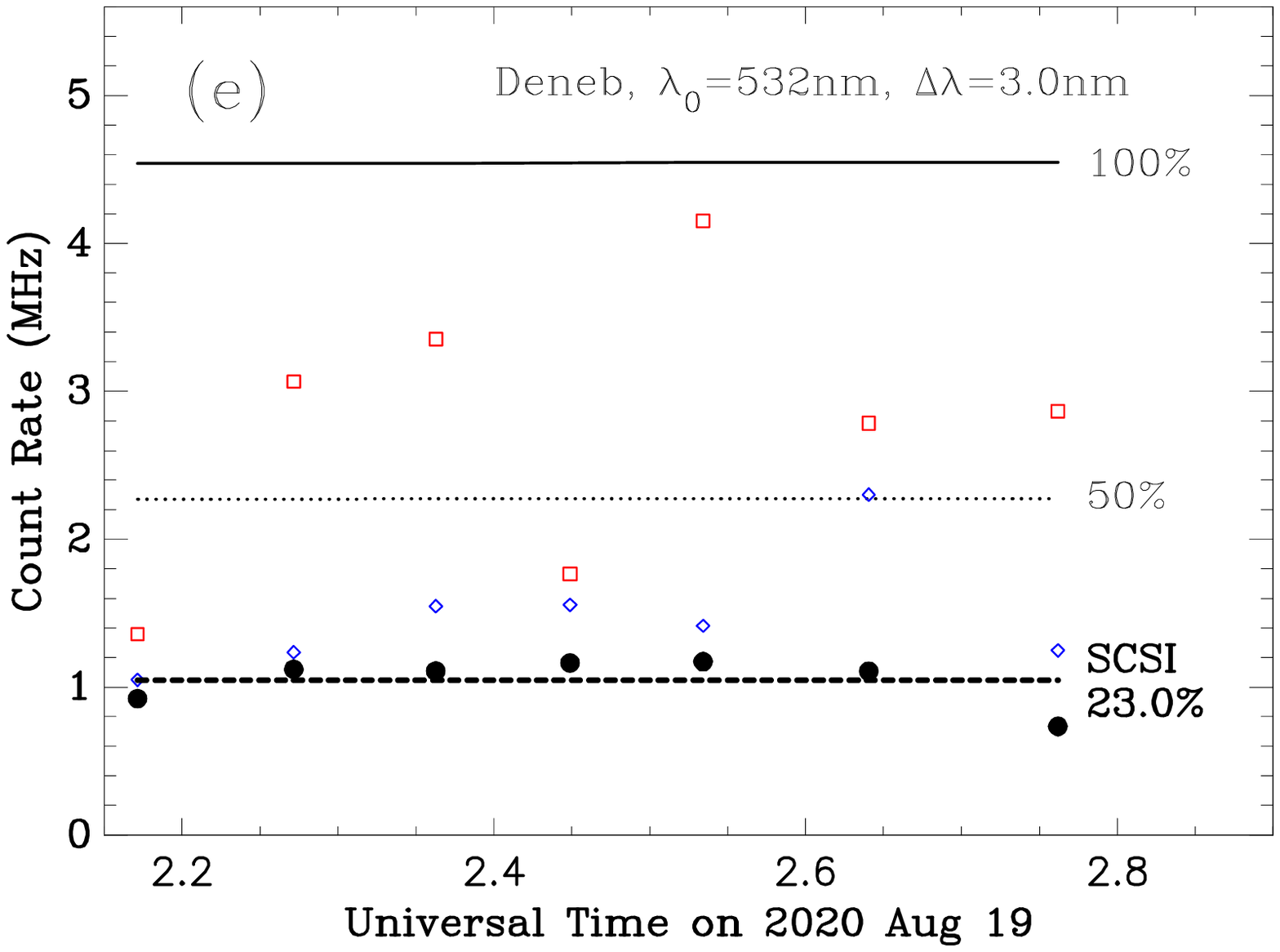}{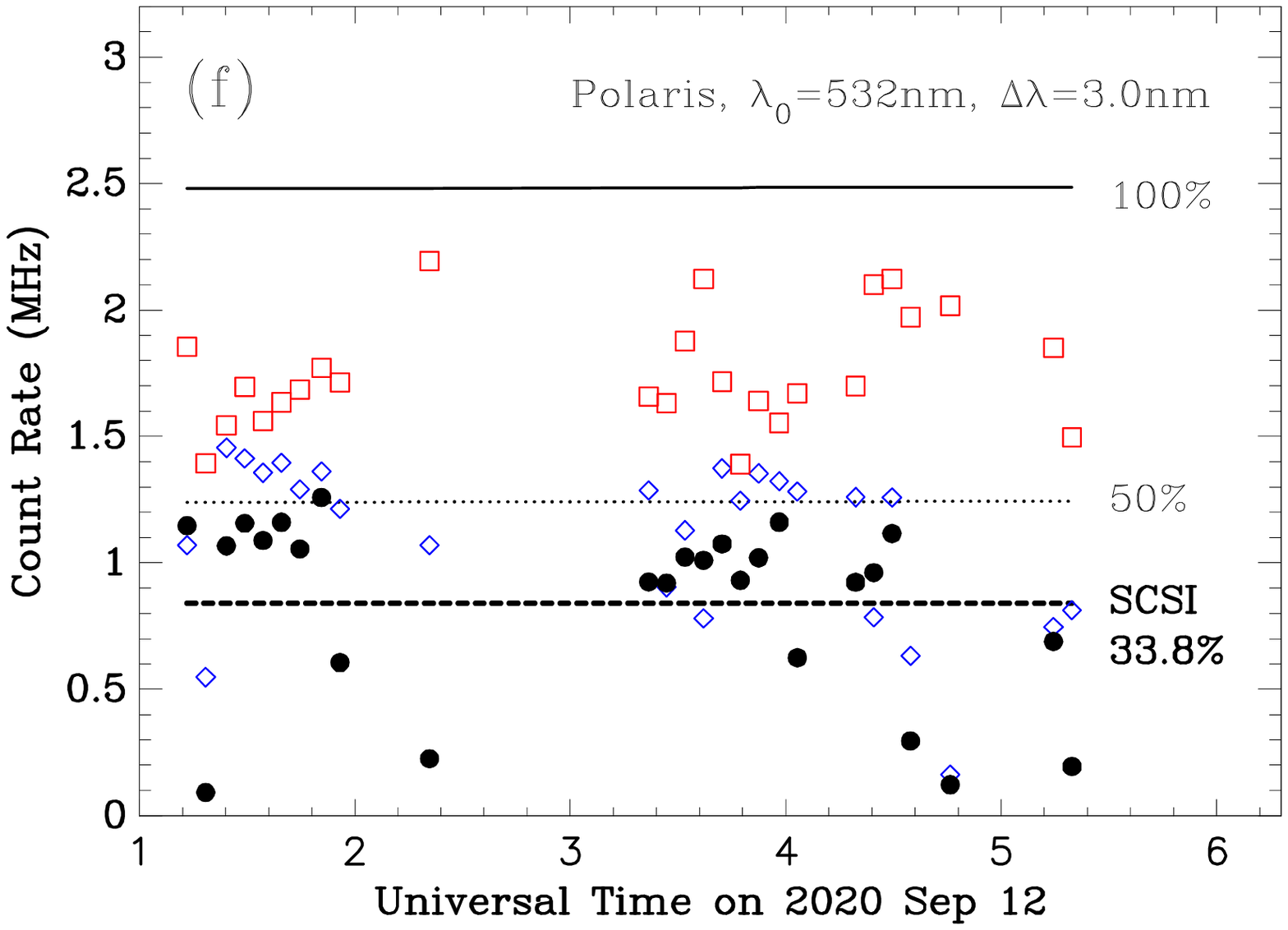}
\figcaption{Six examples of count rates obtained for the full sequence of data files
using the three different filters and observing 
four different stars. In all panels, curves are drawn at 100\% and 50\% of the expected maximum
count rate determined from atmospheric and instrumental parameters as described in the text. 
The blue diamonds and red squares represent
the highest average count rate sustained for a 5 second interval in each file for Channels 0 and 1 
respectively, and the filled circles represent the average count rate obtained for each file as a 
whole (including averaging over both telescopes). 
(a) Vega observed with the wider 532-nm filter. (b) Vega observed with the
633-nm filter. (c) Vega observed with the narrower 532-nm filter. (d) Arcturus observed with the
narrower 532-nm filter. (e) Deneb ($\alpha$ Cyg) observed with the wider 532-nm filter. (f) Polaris observed with
the wider 532-nm filter.}
\end{figure}

A second point can be made using the count rates shown in Figure 6. For the observation
of Vega, the star was rising toward the zenith, and we can see that the telescopes do not track well
enough to keep the star on the detector throughout the file. 
 If we compare
the average count rate that we obtain in a file versus the maximum instantaneous rate, there is as much
as a factor of two difference. 
In contrast, with Polaris, we see that the count rates
are much more consistent, due to the fact that the star is nearly fixed on the sky, hence the 
tracking errors develop more slowly over time and it is easier to stay on the target.
This indicates that a significant increase in average count rate
is possible once better tracking can be made routine.

Figure 7 shows aggregate results for count rates obtained for targets observed on 
6 of the 15 nights listed in 
Table 2. We compare the values recorded with an estimate of the number of photons that
should reach the detectors, given the source brightness, telescope aperture size, 
optical efficiency (including filter width, optical transmission, and atmospheric transmission), 
and detector quantum efficiency.
In each of the six graphs, the expected maximum count rate is shown as the solid black 
curve. This changes over the course of the night because we incorporate into our calculations
the decrease in count rate expected at higher zenith angles. We then fit our count rate data
to a similar curve, and derive an overall efficiency estimate for each observation. Typical 
efficiencies at present are in the 20\% range, with some variation due to the effects discussed
above.

\subsection{Lowell Observations}

Observations at Lowell Observatory occurred on three nights, June 6-8, 2015. The observing
procedure was largely the same as described above except that, because the telescopes 
at Lowell are fixed, the baseline did not change from night to night.  The observations at Lowell were 
also at a much larger separation than we have used in the case of the campus telescopes so far. 
We used the Hall 1.0-m and Perkins 1.8-m telescopes to form the interferometer in this case,
which are separated by 53 m on the Anderson Mesa site. 
The telescope positions are well known, so the issue of placement and its effect on the 
timing precision is removed. Likewise,
telescope pointing and tracking was much less of an issue at Lowell. 
Detector mounts were constructed for the SPADs for each telescope, but these did not include
the two achromats; thus, the filter was not in the collimated beam, leading to a lack of good focus
at the detector. Nonetheless, the apertures were large enough to generate count rates for 
Vega that were comparable to what we achieve with our campus telescopes, roughly 1 MHz
per channel. Over the three nights, we obtained about 5 hours of data on Vega in the 
wider 532-nm filter. The correlator was placed between the two domes and approximately 50
m of cable was stretched from that location to reach the SPAD attached to each telescope.

\section{Analysis of the Data}

\subsection{Data Reduction}

To measure the significance of a correlation peak, one would like to determine the total number 
of correlated events and to compare that with the standard deviation of the number of random 
correlations detected in the same time interval. There are multiple ways that this might be done,
but our approach so far has been to use a binning interval of 64 ps to retain as much signal
as possible while keeping the data processing simple and straight-forward. In the future, 
we hope to continue to improve our reduction tools.

Specifically, we start with a given data file containing the arrival times of photons detected and 
divide the data into ``frames,'' that is, smaller intervals that are determined by the rollover time
in the internal clock of the timing correlator. 
Within each frame, we compute timing differences between events in Channel 2 versus Channel 1. 
These timing differences then populate a 
histogram of timing differences, which is equivalent to the cross-correlation of the two data 
streams. Using the timing information of when the file started and ended, we compute the location
of the star on the sky at the start of the file, and the shift of its position through the file. This determines 
the expected location of the correlation peak in the cross-correlation and its smear
during the file. Using the latter, we shift cross-correlations from each frame accordingly before
coadding them. Finally, a global shift of the cross-correlation is performed to move the starting offset
to a fiducial location (usually zero timing delay). Included in the global shift are the known
intrinsic channel timing differences as well as the cable differences measured in the laboratory. 
Repeating these steps for each file in the observing
sequence, we can coadd to have a final result for the night on that object. Finally, once all 
unresolved targets observed on a given night are analyzed in the same way, we establish the
baseline azimuth offset from north-south by computing curves similar to simulated result shown
in Figure 5. We then boxcar smooth the curve with a width of 0.14 degrees and find the 
local maximum closest to zero offset in the smoothed version. This gives the final azimuth assumed
for a given night.

\subsection{Correlation Results}

If an intensity interferometry
experiment is successful, then the cross-correlation peak should build up with the inclusion
of more and more data files. Rewriting Equation 8 in a slightly different way, we obtain
\beq
{\rm SNR}(T) = \frac{\Delta \tau}{2 \Delta t} \sqrt{\frac{(r \Delta t)^{2} T}{\Delta t}}.
\eeq
The right hand consists of two terms, the first of which is the factor multiplying $|V_{12}|^{2}$
in Equation 4; that is, it represents the fraction of correlated photons. Recalling that $(r \Delta t)^{2}$ 
correlations are produced in a given timing bin of width $\Delta t$, the term under the square root 
may be seen as the total number of correlations obtained in that timing bin for an experiment of duration $T$.
Thus, the SNR is a linear function of the square root of the number of correlations seen in a typical
timing bin in the final cross-correlation function for an observing session, 
where the slope is the fraction of correlated photons.
For examaple, a 1-MHz detection rate at each telescope generates $10^{12}$ correlations per second, 
but these are spread over a large number of 64-ps bins ($\Delta t = 64$ ps). Therefore, a typical
bin will have 64 counts after cross-correlating 1 second of data.  However, if the observing session lasts 
two hours, then 
the cross-correlation function will have an average of 460,800 counts per 64-ps bin. Given a correlation
fraction of 0.24\%, the predicted SNR would be $0.0024 \cdot \sqrt{460800} = 1.63$.
Following this thinking, we can plot the SNR achieved for the observing 
sequences shown in Table 2. This is shown in Figure 8 for all sources expected to be unresolved
at the baselines of our observations. (That is, we do not include observations of
Arcturus here because with a known diameter of 21 mas, the star is expected to be partially
resolved and to have a reduced correlation.) The plot indicates that the SNRs obtained are
generally in the expected range, with the longest observing
generating the highest SNRs. We plot these SNR values as a function of the square-root of the
mean number of correlations per 64-ps bin; this should be a linear relationship with SNR, with
the slope being the correlation fraction. 
Taking the data at face value and fitting a line to the complete data
set regardless of filter, we obtain a slope of 
$0.00231 \pm 0.00039$.  The wider of the two 532-nm filters dominates the data set at
present, so we find our 
correlation fraction in line with the expected value discussed in Section 2.

\begin{figure}[t]
\figurenum{8}
\hspace{3.5cm}
\includegraphics[scale=0.60]{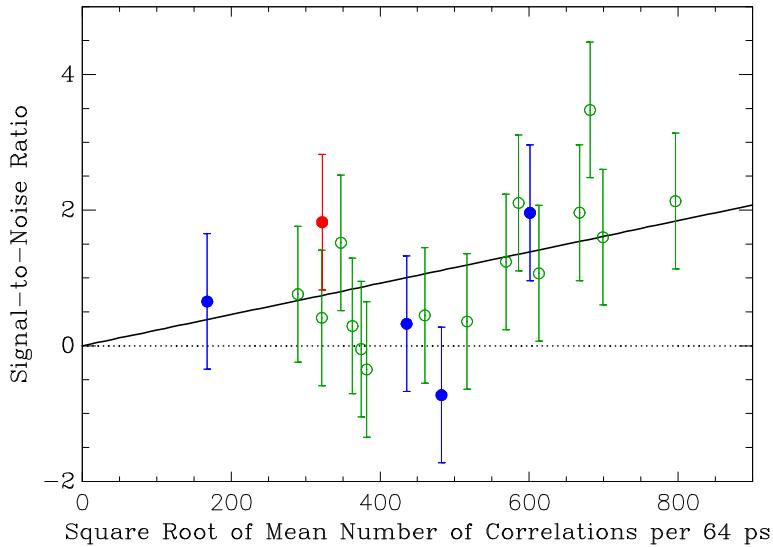}
\caption{
The signal-to-noise ratio of the observations shown in Table 2 as a function of the square root
of the mean number of correlations per 64-ps interval. The open circles represent observations
taken with the 532/3-nm filter, whereas filled circles represent the two narrower filters, with blue for
the 532/1.2-nm and red for the 633/1-nm filter. The linear fit obtained on the entire set of observations
is shown as a solid line. The dotted line at zero is meant to guide the eye.}
\end{figure}

Taking this same set of observations, we can coadd all of the results to generate a single cross-correlation plot. Although this contains data of four separate sources, all are expected to be
unresolved from previous diameter determinations. This is shown in Figure 9 in four different
representations. Panels (a) and (b) show the cross-correlation result at a resolution of 16 ps.
We do not measure the SNR from these plots, but the cross-correlation peak should
have a width which is close to 48 ps according to the detecter parameters we have measured.
It can be seen in panel (a) that, if the baseline azimuth tuning is not done in the data reduction,
there is a spreading out of the peak, whereas when the tuning is applied, the result is that in 
panel (b). Here, the peak does have the expected width of  3-4 timing samples, as well as the 
correct position, once all of the timing offsets have been applied. In Figures 9(c) and (d), 
we show the same data as (a) and (b) respectively, but at a timing resolution of 64 ps. In this
case, the correlation should appear as essentially a delta function. The SNR can be estimated
by comparing the number of excess correlations in that single sample point with the samples
nearby. For the noise calculation, we use the samples between -2 and 2 ns, 
excluding the three central samples, the middle of which has the correlation peak. 
We obtain a SNR of 4.70 if the
baseline azimuths are not tuned in software, and 6.76 when that tuning is applied.

\begin{figure}[!t]
\figurenum{9}
\plottwo{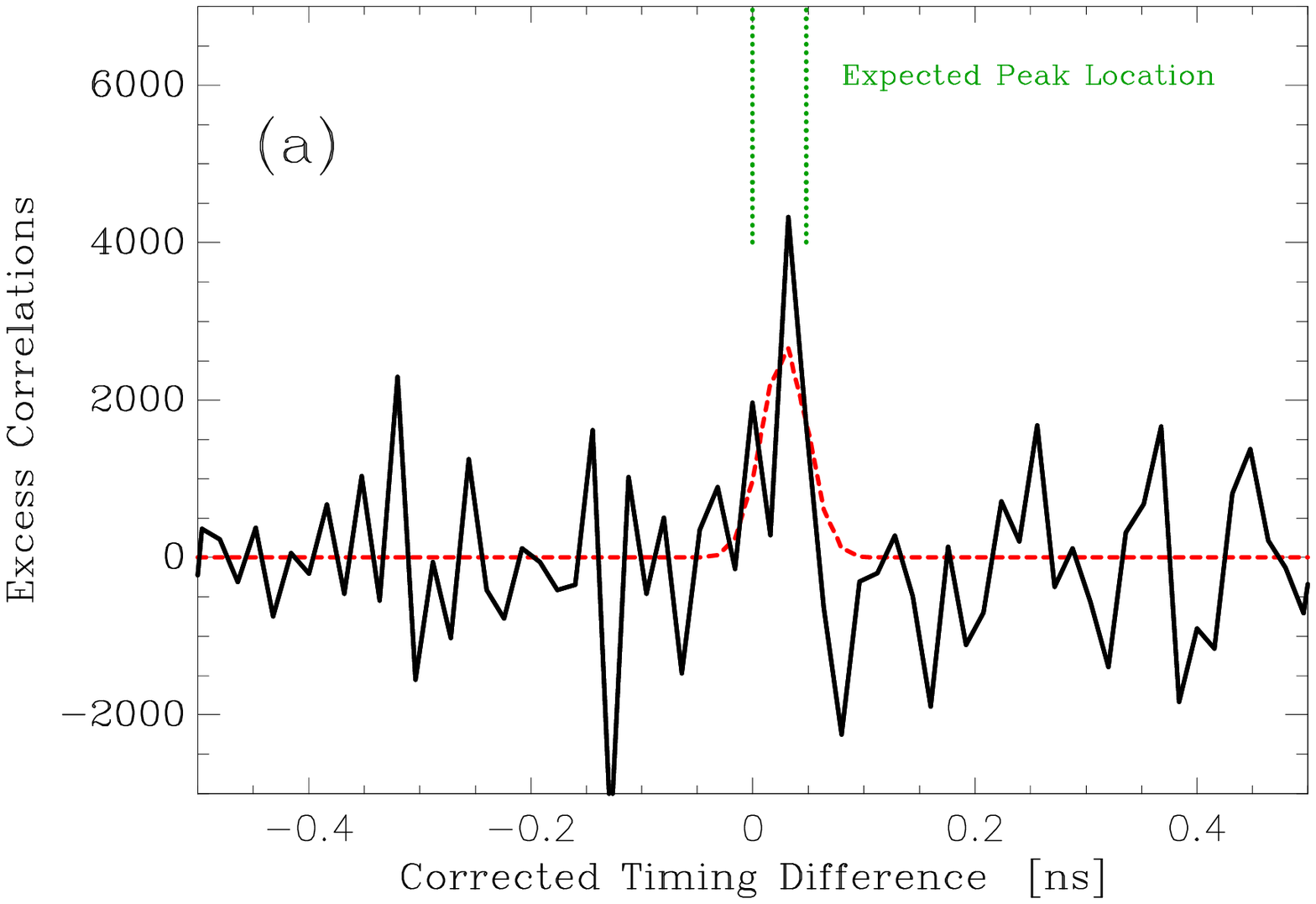}{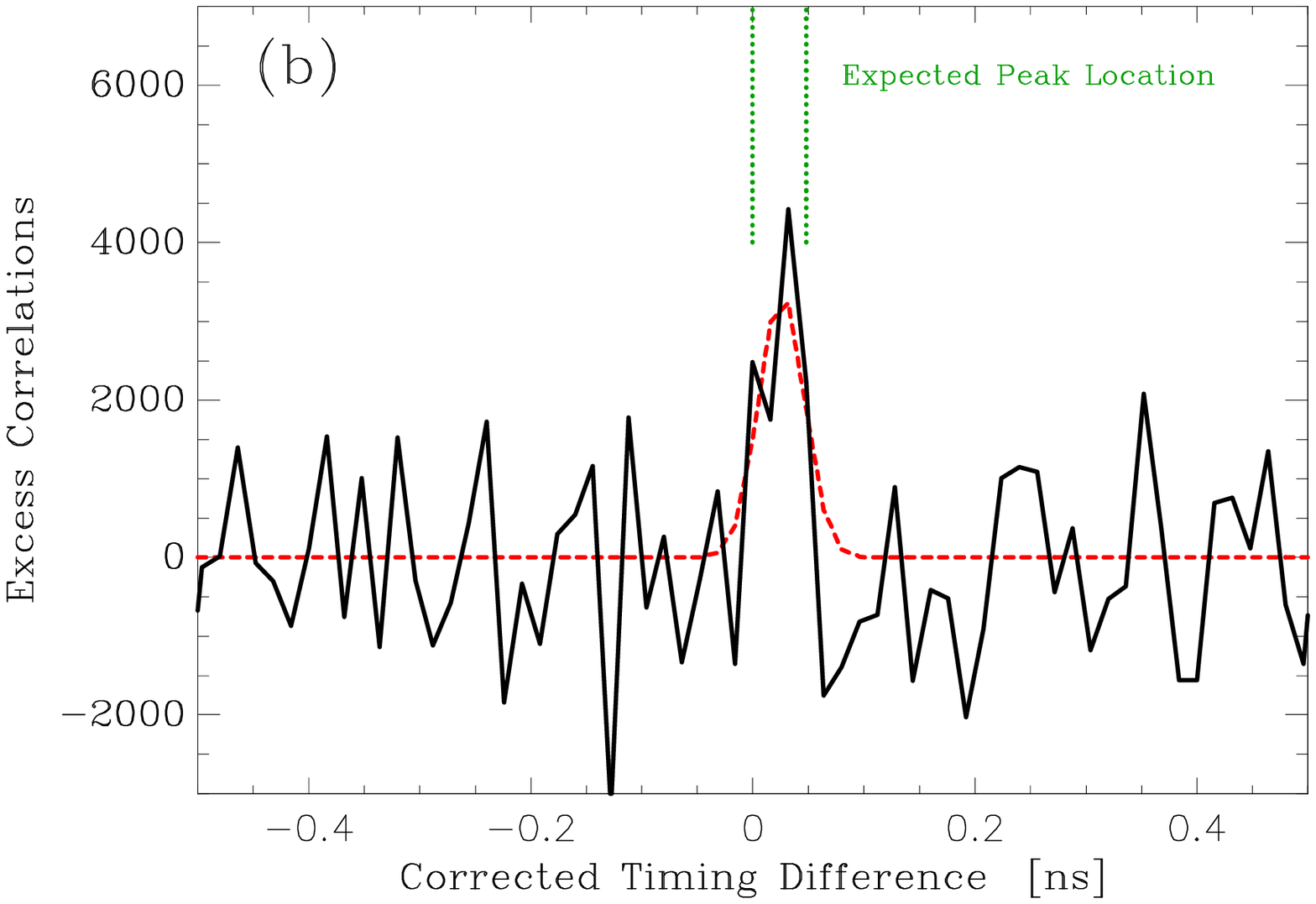}

\plottwo{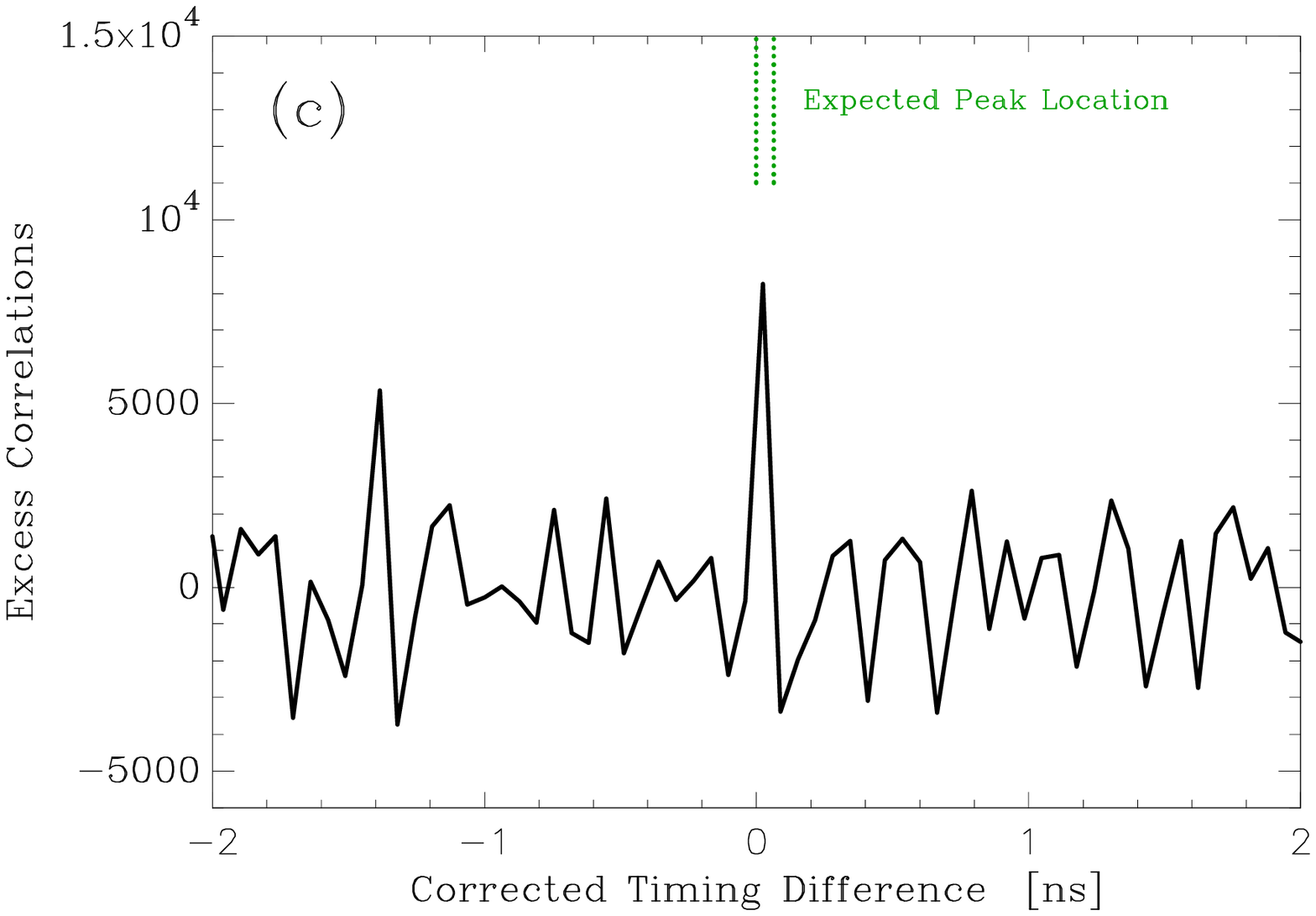}{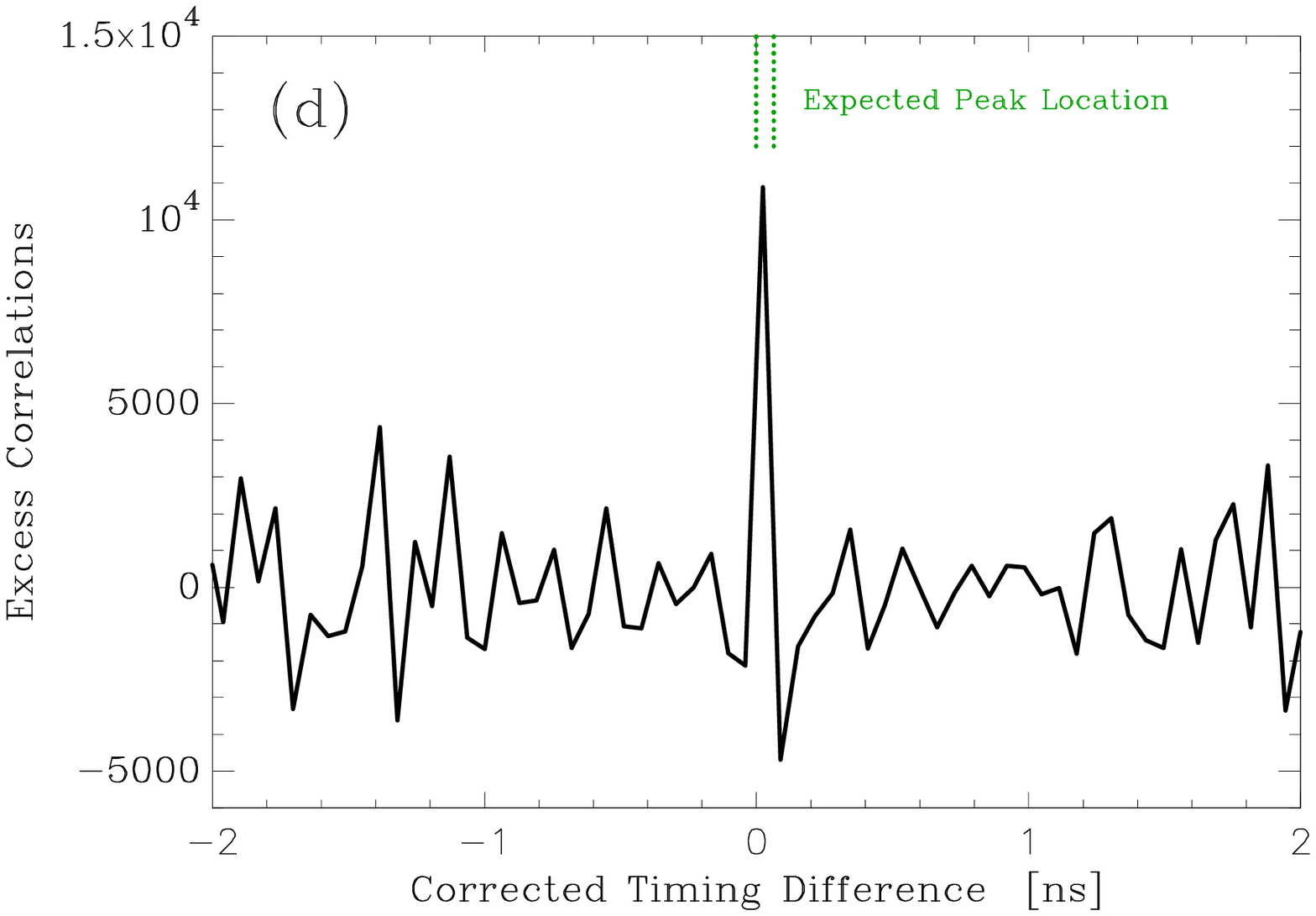}
\figcaption{Excess correlations obtained as a function of corrected timing difference when 
summing all observations where the source is expected to be unresolved. (a) In this plot,
no correction for the baseline azimuth has been applied. (b) The same data as in (a), but in
this case the baseline azimuth tuning described in the text has been applied for each night
prior to summing the results. In these two panels, the plots are drawn at a resolution of 16 ps
and the green vertical lines indicate the expected placement of the peak after the
timing delays are corrected to the same fiducial. The red dashed curve indicates the 
best fit Gaussian of FWHM of 48 ps centered on the predicted location. 
(c) The same data as in (a), but drawn at
a resolution of 64 ps. The formal significance of the peak is = 4.70$\sigma$ above the noise. 
(d) the same data as for (b), but at a resolution of 64 ps. The formal significance here
is 6.76$\sigma$. In (c) and (d), the expected location of the peak is again marked.}
\end{figure}

Finally, we took these data and separated out the measures of each star to create five 
individual excess correlation plots at the final timing resolution of 
64 ps (i.e., the same as in Figures 9[c] and [d]). Setting aside our single observation of 
Altair ($\alpha$ Aql), Figure 10 shows the results for the four other stars we have observed: Deneb, Polaris,
Arcturus, and Vega. The fifth plot is again of Vega, showing the result from Lowell observations 
only.
Given these results, we would expect that our SCSU observations of Vega, Deneb, and 
Polaris should show a correlation peak, as our observations were taken at baselines
generally under 3 m on campus. (Polaris is the exception, where, at a baseline of 9 m the
correlation peak may start to decrease.) In contrast, Arcturus has a much larger angular
diameter, and the Lowell observations of Vega are at baselines large enough to expect
little or no correlation. Although the SNR is low in the plots shown in Figure 10, this trend 
is indeed seen, where the highest correlations are seen for Deneb and the SCSU observations
of Vega, whereas the results for Polaris and Arcturus are consistent with expectations of
partial correlation within the uncertainties, and there is
no evidence of correlation with the Lowell observations of Vega.

\begin{figure}[t]
\figurenum{10}
\includegraphics[scale=0.33]{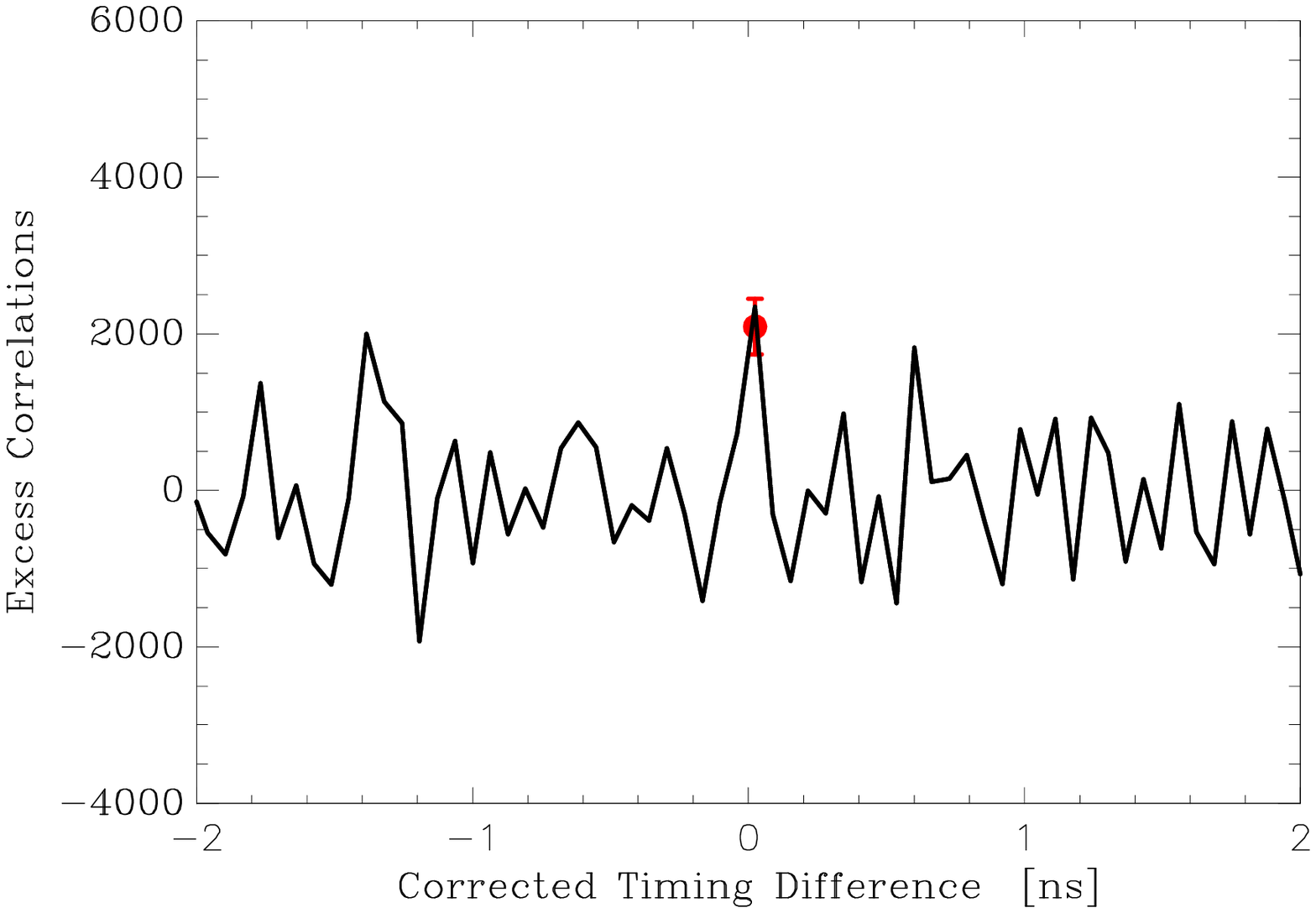}
\includegraphics[scale=0.33]{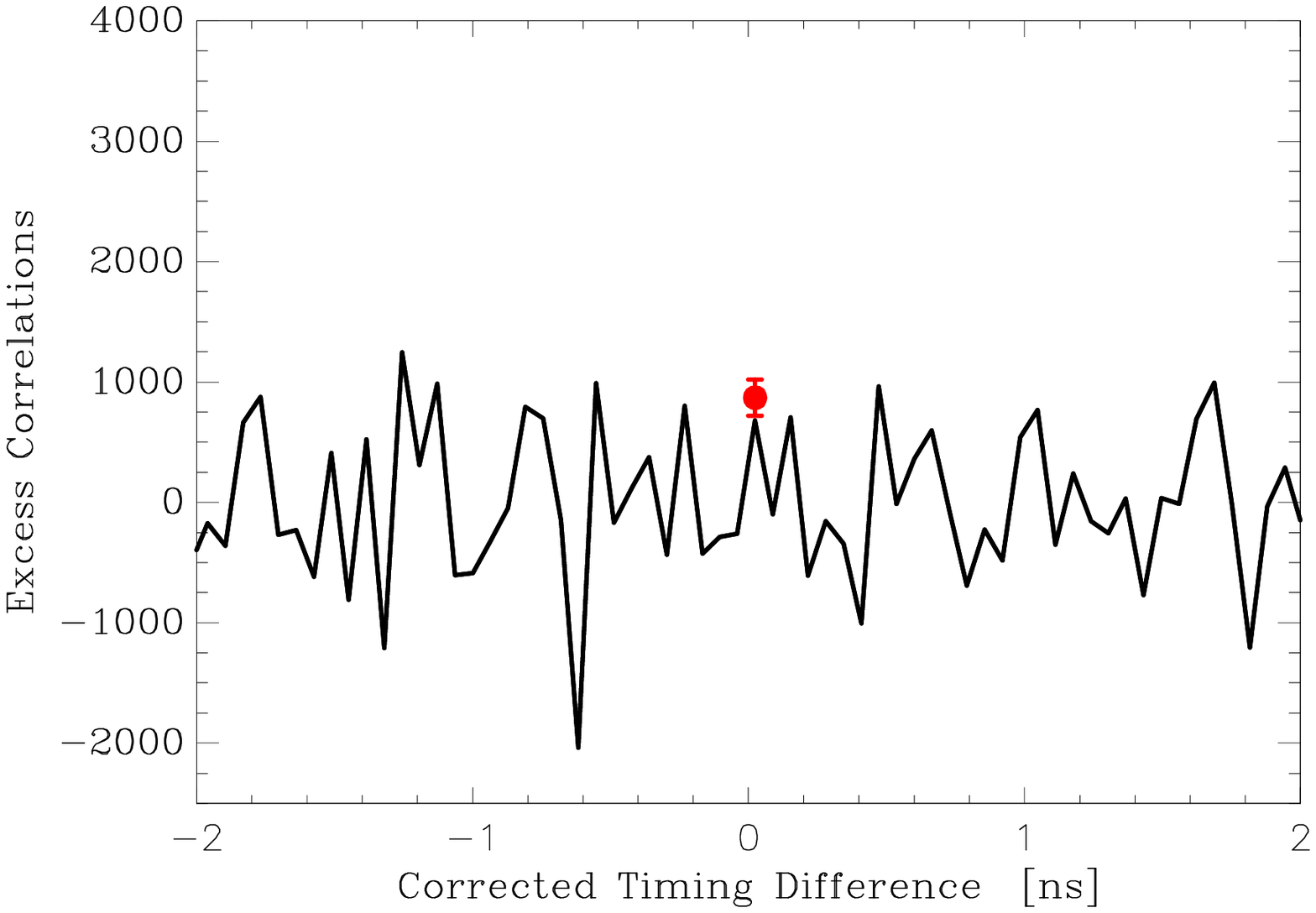}
\includegraphics[scale=0.33]{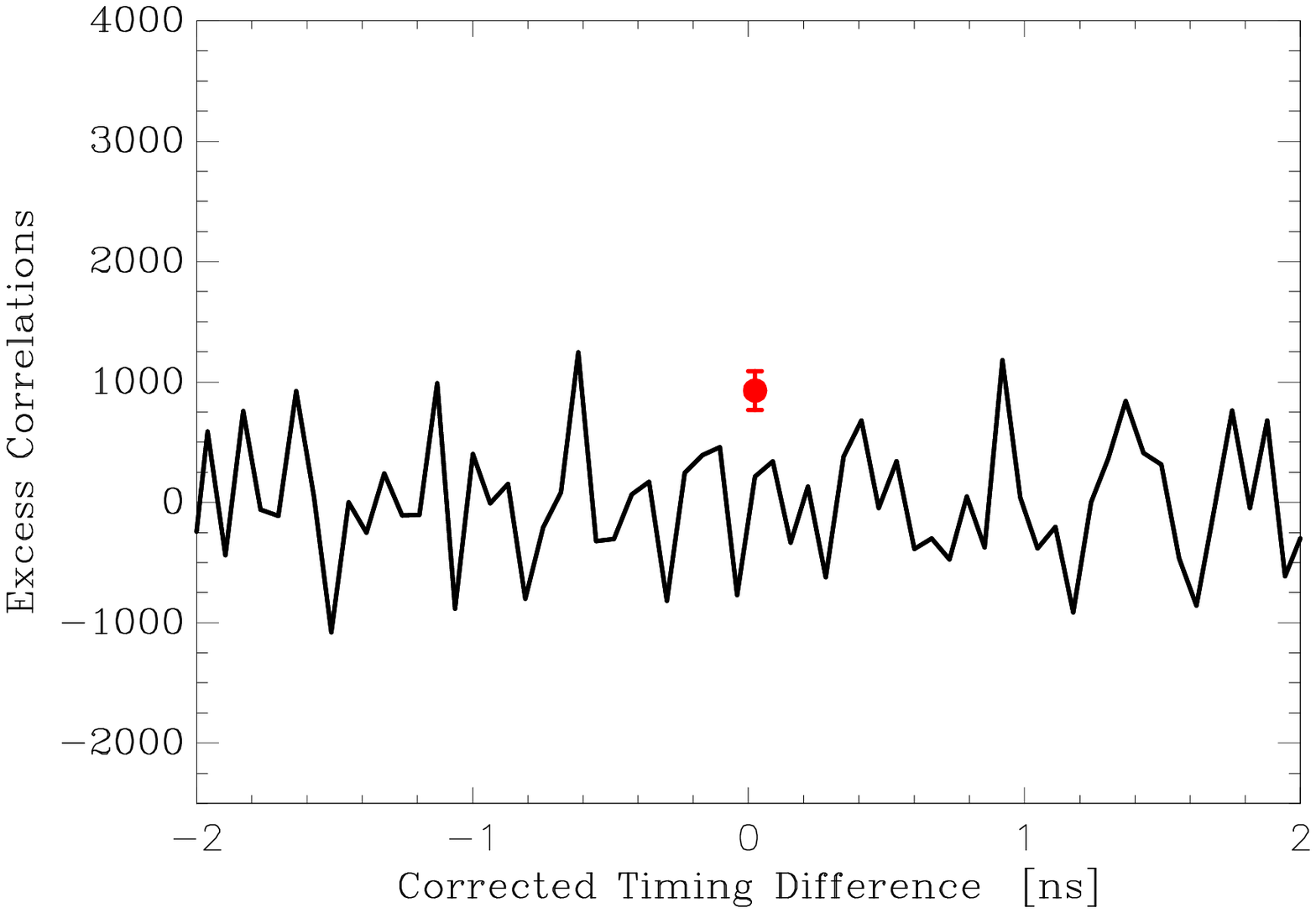}

\vspace{0.5cm}

\hspace{3.25cm}
\includegraphics[scale=0.33]{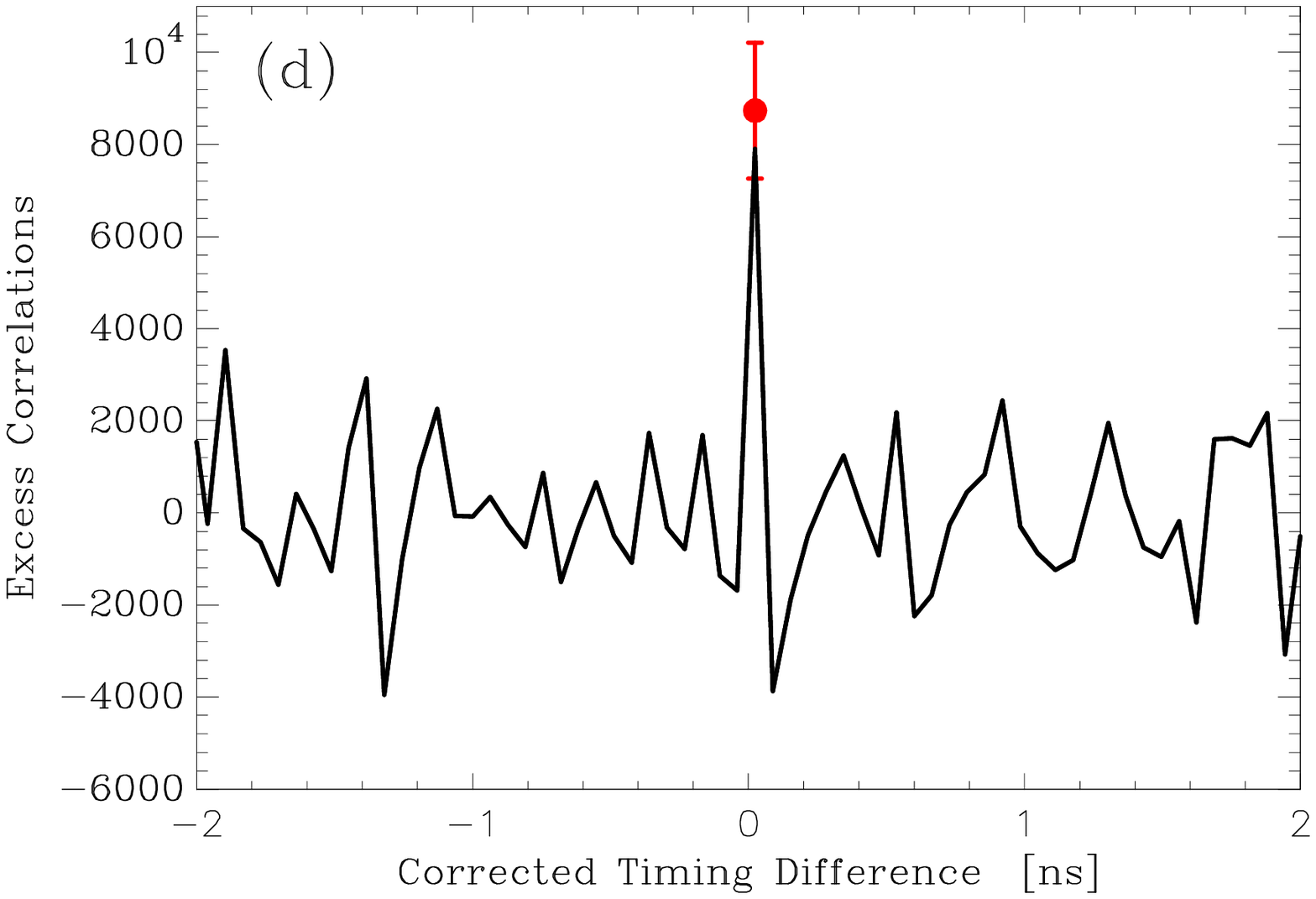}
\includegraphics[scale=0.33]{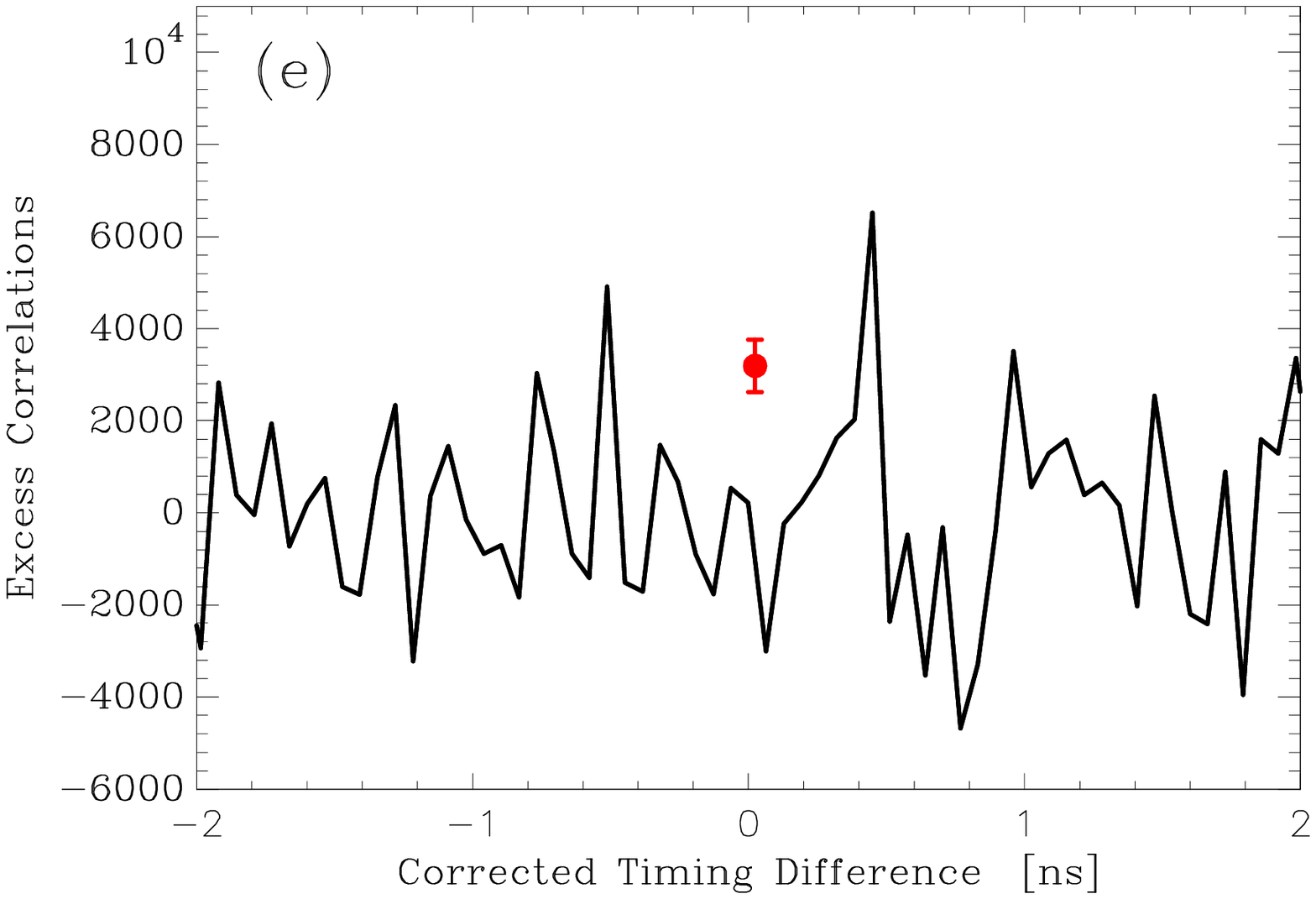}

\caption{
Individual excess correlation plots for each star. (a) Deneb. (b) Polaris. (c) Arcturus. (d) Vega 
observed at the
smaller baselines used at SCSU. (e) Vega observed at the larger baselines at Lowell. 
In all plots, the timing resolution is 64 ps and the red filled circle represents the 
predicted location of a full-correlation peak given the mean number of correlations
of each data set and 
assuming a correlation fraction of $0.000231 \pm 0.00039$.}
\end{figure}

\begin{deluxetable}{llrrcll}
\tablewidth{0pt}
\tablenum{3}
\tablecaption{Previous Angular Diameter Measurements for Star Observed with SCSI}
\tablehead{
\colhead{Object} & Bayer & HR & HD &
\colhead{$\theta_{LD}$}  & \colhead{Instrument} &
\colhead{Reference} \\
& Desig. &&& \colhead{(mas)} &  & 
}
\startdata
Altair & $\alpha$ Aql & 7557 & 187642 & $3.462 \pm 0.035$ & Mark III & \citet{moz03} \\
Altair &  $\alpha$ Aql & 7557 & 187642 & $3.309 \pm 0.006$ & NPOI & \citet{bai18} \\
Arcturus & $\alpha$ Boo & 5340 & 124897 & $21.373 \pm 0.247$ & Mark III & \citet{moz03} \\
Deneb & $\alpha$ Cyg & 7924 & 197345 & $2.420 \pm 0.060$ & Mark III & \citet{moz03} \\
Polaris & $\alpha$ UMi & 424 & 8890 & $3.123 \pm 0.008$ & CHARA & \citet{mer06} \\
Vega & $\alpha$ Lyr & 7001 & 172167 & $3.24 \pm 0.07$ & Narrabri & \citet{han74} \\
Vega & $\alpha$ Lyr & 7001 & 172167 & $3.28 \pm 0.01$ & PTI\tablenotemark{a} & \citet{cia01} \\
Vega & $\alpha$ Lyr & 7001 & 172167 & $3.225 \pm 0.032$ & Mark III & \citet{moz03} \\
Vega & $\alpha$ Lyr & 7001 & 172167 & $3.280 \pm 0.006$ & NPOI & \citet{bai18} \\
Vega & $\alpha$ Lyr & 7001 & 172167 & $3.33 \pm 0.01$ & CHARA & \citet{auf06} \\
\enddata
\tablenotetext{a}{Palomar Testbed Interferometer, \citet{col99}.}
\end{deluxetable}

\subsection{Visibility Measurements}

Given that the instrument is detecting excess photon correlations at the expected 
level, we seek to use Equation 4 to derive information about the visibility as a function of 
baseline for our observations, in order to make statements about the diameters of the stars
observed. For comparison, we show a summary of previous diameter measurements of these 
four stars in Table 3.

To begin this process, we return to \citet{dyc99}, where the relationship between the object's
irradiance distribution on the sky is related to its two-dimensional Fourier transform. 
Defining $u$ and $v$ to be the Fourier conjugate variables to orthogonal spatial
coordinates $x$ and $y$ on the image plane, the $(u,v)$ components
in the Fourier plane are determined by the following equations involving the observational 
parameters discussed in Section 4.3:

\beq
u = (-b_{N} \sin \phi \sin h + b_{E} \cos h + b_{Z} \cos \phi \sin h) / (206265 \lambda)
\eeq

\noindent
and 

\beq
v = [b_{N}(\sin \phi \sin \delta \cos h + \cos \phi \cos \delta) + b_{E} \sin \delta \sin h - 
b_{Z}(\cos \phi \sin \delta \cos h - \sin \phi \cos \delta) ] / (206265 \lambda).
\eeq

\noindent
We can use the above to generate a $(u,v)$ plot of observations on a given target each time
we observe it. Four examples are shown in Figure 11. We note that, although these sequences
trace out an arc in the $(u,v)$-plane, the magnitude of the spatial frequency sampled 
generally does not change very much. For radially symmetric sources observed over the 
period of time and sky position that we typically observe them, we 
are normally acquiring signal at a very narrow range of the magnitude of the spatial
frequency vector, $|{\bf u}|$.

This allows us to combine the data on each star and plot it as a function of $|{\bf u}|$. We
normalize $|{\bf u}|$ in the typical way as

\beq
|{\bf u}| = \frac{b}{206265 \lambda},
\eeq

\noindent
where the units of $|{\bf u}|$ will be cycles per arcsecond. However, to put all our observations
on equal footing given the known angular diameters, we can parameterize the relationship 
between the spatial frequency sampled and the width of the Airy disk in the Fourier plane
using the following expression:

\beq
x = \pi \theta_{\rm LD} |{\bf u}| = \frac{\pi b \theta_{\rm LD}}{206265 \lambda}.
\eeq

\noindent
Here, following \citet{bai18}, $ \theta_{\rm LD}$ represents the limb-darkened angular diameter
of the (radially symmetric) source. The advantage of this parameterization is that, regardless
of a star's angular diameter, the first zero of stellar profile in the Fourier domain will occur
in the same place (which would be $x=3.83$ for a perfect Airy pattern, the Fourier transform 
of a uniform disk). In \citet{bai18}, those authors use a limb darkening model which depends
on a parameter they define as $\mu$, which they measure for each star that they report on. 
The range of 
$\mu$ varies from approximately 0.4 to 0.8. In Figure 12(a), we show limb-darkened profiles
for these two extremes of $\mu$ using the same formulation as in \citet{bai18}, 
and overplot our results for the four stars for which we 
can measure the squared visibility (the same stars as in Figure 10). These show that as the 
baseline approaches the first zero of the parameterized stellar profile, the correlation we 
observe decreases. From this, we conclude that it is possible to measure stellar 
diameters with our instrument.

In Figure 12(b), we replot the results obtained so far for Vega and Arcturus, where
in this case the $x$-coordinate is plotted in cycles per arcsecond. We then change the value
of $\theta_{\rm LD}$ so that the visibility curves pass through the upper limit of the error bar for the measurement
at highest spatial frequency. For Vega, this data point is the one derived from the observations
at Lowell, and we can pass a second pair of curves through
the lower limit of the small-$|{\bf u}|$ 
observation in a similar fashion (this data point being derived from the on-campus observations). 
The values of $\theta_{\rm LD}$ corresponding to these
curves then set limits on the range of the
diameters for each star that are currently consistent with our data. For Arcturus, we find 
$\theta_{\rm LD} > 15$ mas, and for Vega, $17$ mas $> \theta_{\rm LD} > 0.8$ mas. 
While the uncertainties are large, these results are nonetheless consistent with the previous 
measures shown in Table 3.

Our results for Vega may be directly compared with those of a recent experiment observing the same
star with the Asiago 
Stellar Intensity Interferometer \citep{zam21}. That instrument has a much larger distance between
the two telescopes used than at Lowell, but the telescopes used in both cases are comparable in size and the 
detection and timing instrumentation likewise has similar properties overall. Both their results and ours
show a high degree of consistency, giving confidence that the SPAD approach with small telescopes is 
a viable means with which to conduct intensity interferometry.

\begin{figure}[t]
\figurenum{11}
\hspace{3.5cm}
\includegraphics[scale=0.60]{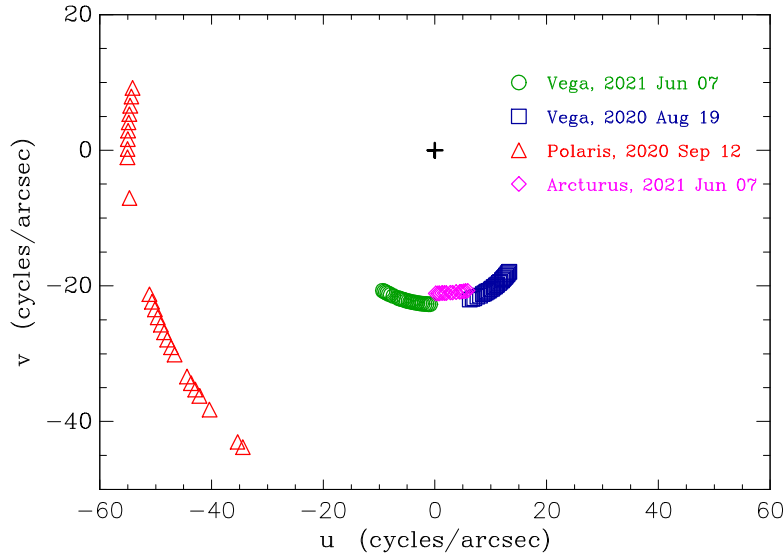}
\caption{
Fourier plane coverage for four representative observing sequences.}
\end{figure}

\begin{figure}[!t]
\figurenum{12}
\plottwo{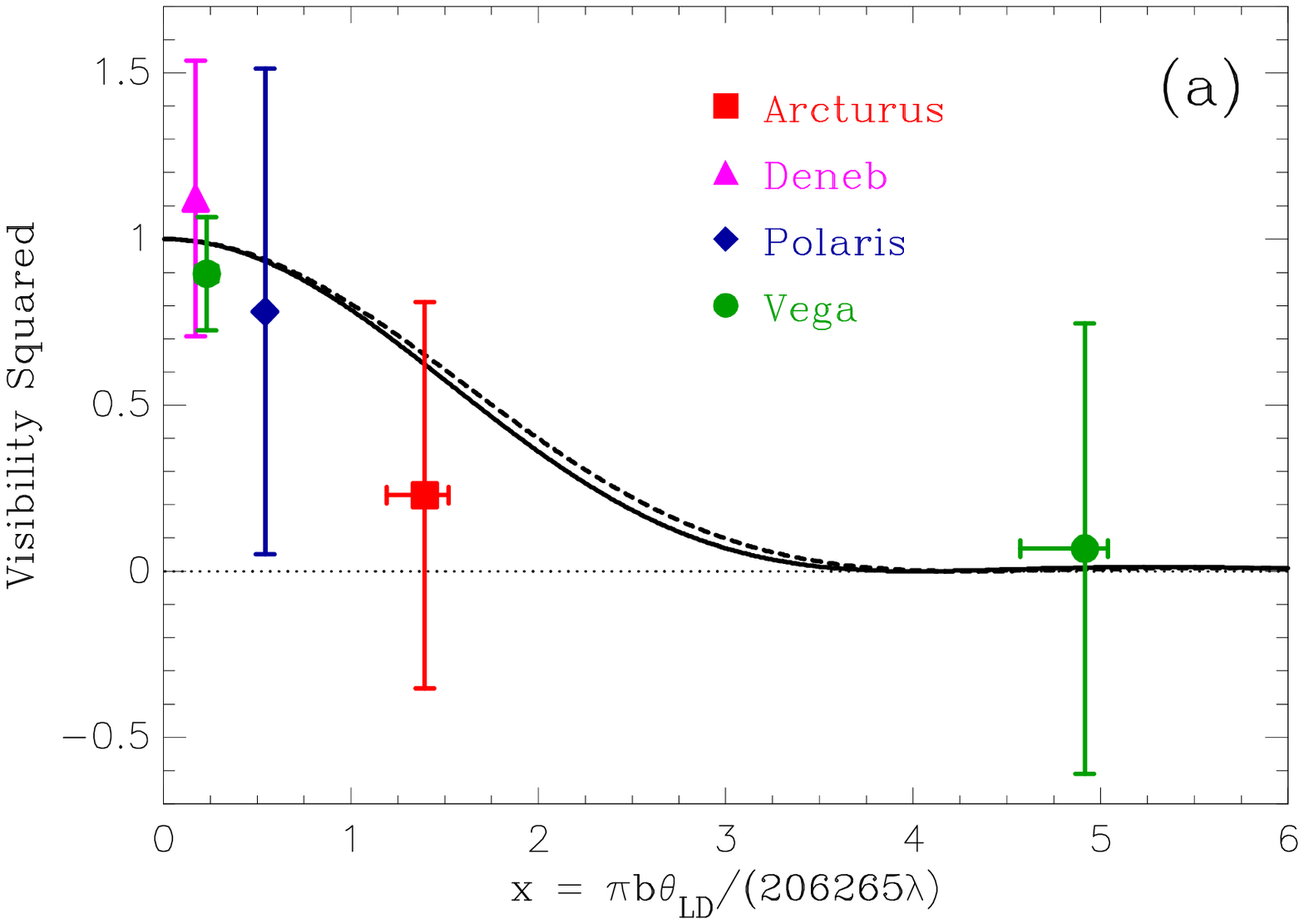}{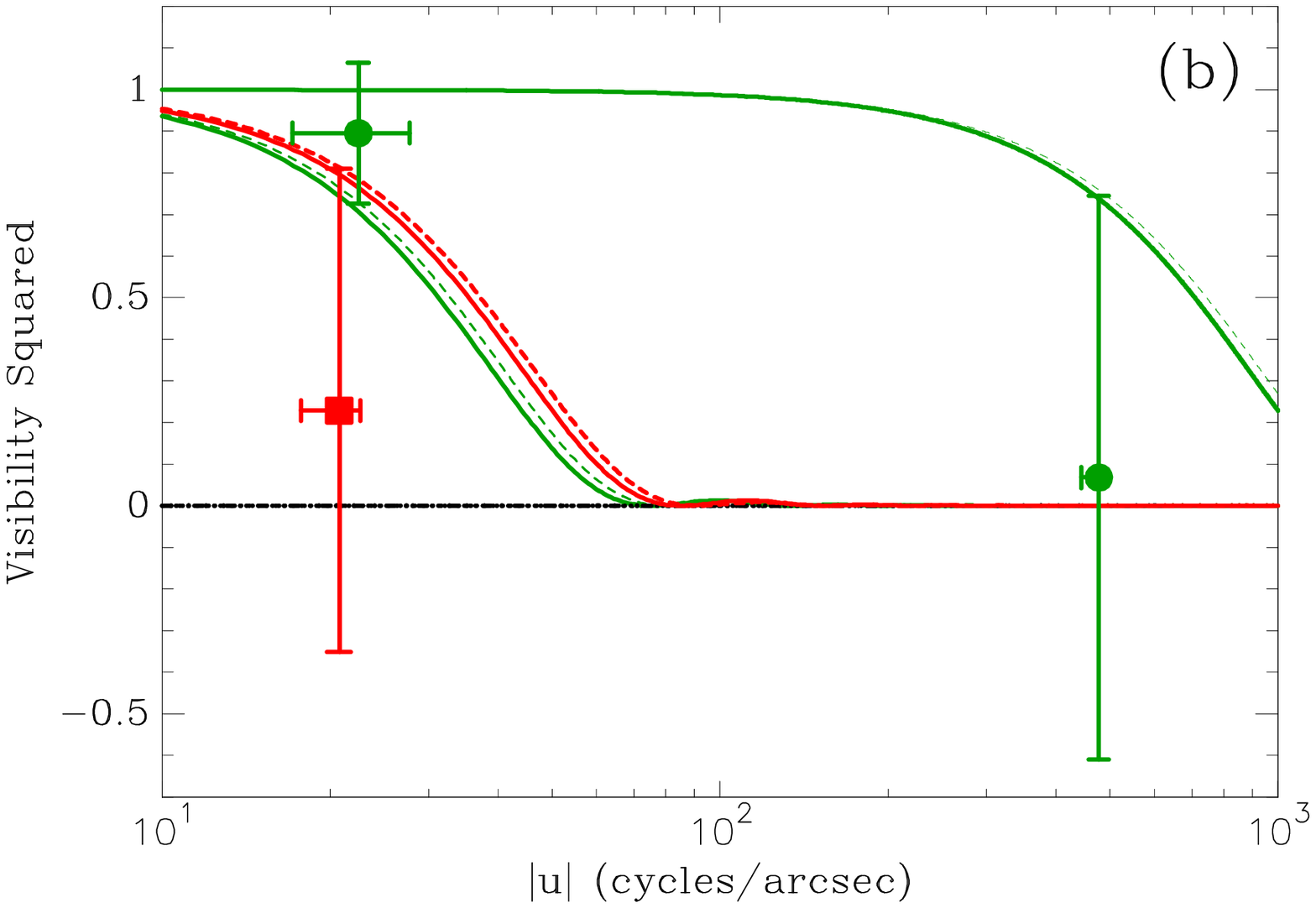}
\figcaption{
(a) Visibility plot for observations taken to date with SCSI. The solid and dashed curves 
indicate the two limb-darkened profiles of different $\mu$ value as discussed in the text. 
(b) The points for Vega and
Arcturus are plotted again with the same plot symbol and color, 
but as a function of the spatial frequency. The solid and 
dashed lines are the limb-darkened profiles that represent the 1-$\sigma$ lower limit of the 
angular diameter as discussed for Arcturus and both upper and lower limits for Vega. 
The abscissa in (b) is plotted on a log scale for clarity.}
\end{figure}

\section{Conclusions and Future Work}

We have built a new stellar intensity interferometer which is based on detecting coincident
photons at two or more different telescopes using a high-precision timing module and single photon
avalanche diode detectors. The timing precision of the instrument is found to be approximately
50 ps. This allows us to use smaller telescopes that have been used in other modern 
intensity interferometers, although it also requires us to measure our baselines and other
instrument parameters very precisely.
We have shown that photon correlations are observed at the 
level expected based on the design of our instrument, and that the data taken to date are 
consistent with partial correlation within the uncertainties
for Arcturus and Polaris. In taking our instrumentation to Lowell Observatory, 
we observed Vega at a baseline of approximately 50 m using the Hall and Perkins telescopes
on Anderon Mesa. This allowed us to add a null result for Vega at that baseline, with large
uncertainty. Together
these observations permit us to conclude that the angular diameter of Arcturus is larger
than 15 mas, and that of Vega between 0.8 and 17 mas. As we have timing correlators with
up to 8 channels, our instrument is easily upgraded to include more telescopes;
we have already integrated a third telescope identical to the original two and plan to 
use all three telescopes in ongoing observations moving forward. This will permit
us to observe at three baselines simultaneously, making our observations much more 
efficient. 

\acknowledgements 

We gratefully acknowledge support from the National Science Foundation, specifically grants 
AST-1429015 and AST-1909582. 
We thank Daniel Nusdeo and Genady Pilyavsky, who helped us obtain the 
observations at Lowell Observatory, and Lawrence Wasserman, Otto Franz, and Gerard van Belle, who also enthusiastically supported our use of Lowell facilities for this project and
helped with various practical aspects of our visit. We are also grateful to William van Altena,
who read the manuscript prior to submission, and the anonymous referee; both made helpful suggestions. 
Finally, we acknowledge William Shakespeare, who 
wrote the words that have guided us through long nights of observing ({\it Henry V}, 
Act IV, Scene III).






\end{document}